\documentclass[11pt]{JHEP3} % 10pt is ignored!

\usepackage{epsfig,multicol,bbm}

\newcommand\fverb{\setbox\fverbbox=\hbox\bgroup\verb}
\newcommand\fverbdo{\egroup\medskip\noindent%
            \fbox{\unhbox\fverbbox}\ }
\newcommand\fverbit{\egroup\item[\fbox{\unhbox\fverbbox}]}
\newbox\fverbbox

\usepackage{amsthm,amscd,dsfont}

\newcommand{\CE}{\mathcal{E}}
\newcommand{\CL}{\mathcal{L}}
\newcommand{\CO}{\mathcal{O}}
\newcommand{\CN}{\mathcal{N}}

\newcommand{\CH}{\mathcal{H}}

\newcommand{\Z}{\mathds{Z}}
\newcommand{\C}{\mathds{C}}
\newcommand{\R}{\mathds{R}}

\newcommand{\nn}{\nonumber}

\newcommand{\spa}{\ , \ \ }

\newcommand{\tr}{\mathop{{\rm Tr}}}

\newcommand{\ads}{\mbox{AdS}}

\newcommand{\sectiono}[1]{\section{#1}\setcounter{equation}{0}}

\title{Finite-size corrections to the rotating
string and the winding state}

\author{Davide Astolfi$^1$, Troels Harmark$^2$, Gianluca Grignani$^1$, Marta Orselli$^2$\\
$^1$Dipartimento di Fisica, Universit\`a di Perugia,\\
I.N.F.N. Sezione di Perugia,\\
Via Pascoli, I-06123 Perugia, Italy\\
    E-mail: \email{astolfi@pg.infn.it}, \email{grignani@pg.infn.it}\\
    \\
$^2$The Niels Bohr Institute  \\
Blegdamsvej 17, 2100 Copenhagen \O , Denmark \\
E-mail:\email{harmark@nbi.dk}, \email{orselli@nbi.dk}}

\abstract{We compute higher order finite size corrections to the energies of the circular rotating
string on $AdS_5\times S^5$, of its orbifolded generalization on $AdS_5\times S^5/\Z_M$ and of the
winding state which is obtained as the limit of the orbifolded circular string solution when
$J\to\infty$ and $J/M^2$ is kept fixed. We solve, at the first order in $\lambda'=\lambda/J^2$, where
$\lambda$ is the 't Hooft coupling, the Bethe equations that describe the anomalous dimensions of the
corresponding gauge dual operators in an expansion in $m/K$, where $m$ is the winding number and $K$
is the ``magnon number", and to all orders in the angular momentum $J$. The solution for the circular
rotating string and for the winding state can be matched to the energy computed from an effective
quantum Landau-Lifshitz model beyond the first order correction in $1/J$. For the leading $1/J$
corrections to the circular rotating string in $m^2$ and $m^4$ and for the subleading $1/J^2$
corrections to the $m^2$ term, we find agreement. For the winding state we match the energy
completely up to, and including, the order $1/J^2$ finite-size corrections.

The solution of the Bethe equations corresponding to the spinning closed string is also provided in
an expansion in $m/K$ and to all orders in $J$.}

\preprint{} 

\begin{document}

%%%%%%%%%%%%%%%%%%%%%%%%%%%%%%%%%%%%%%%%%%%%%%%%%%%%%%%%%%%%%
\sectiono{Introduction}

Semi-classical closed string states on $\ads_5\times S^5$
\cite{Gubser:2002tv,Frolov:2002av,Frolov:2003qc} and their gauge
theory duals, local composite operators of $\mathcal{N}=4$ Super
Yang-Mills (SYM) theory, have recently played an important role in
the understanding of the AdS/CFT correspondence. The discovery of
integrable structures in planar $\mathcal{N}=4$ SYM theory
\cite{Minahan:2002ve,Beisert:2003tq,Beisert:2003yb,Beisert:2003ys}
and tree-level string theory on $AdS_5\times S^5$
\cite{Mandal:2002fs,Bena:2003wd} has sparked the hope of being able
to match the spectrum of semi-classical string states with that of
their dual gauge theory operators.

Considering as an example semi-classical string states with a large
angular momentum $J$ on $S^5$, corresponding to an R-charge in
$\CN=4$ SYM, one can have $\lambda' = \lambda/J^2$ small ($\lambda$
being the 't Hooft coupling of $\CN=4$ SYM theory) on both sides of
the correspondence, in gauge theory by expanding in $\lambda$, and
on the string side by expanding in $1/J$ in the semi-classical
regime $\lambda \gg 1$. While agreement is found at first and second
order in $\lambda'$, for the leading and the first $1/J$ correction,
the agreement breaks down for $\lambda'^3$, a disagreement known as
the three-loop
discrepancy~\cite{Serban:2004jf}.%
\footnote{See also
\cite{Callan:2003xr} for a closely related discrepancies in the near
plane wave/BMN correspondence also cured by the introduction of the
dressing factor.} Recently a substantial effort has been made to
remedy this disagreement, in order to establish an interpolation
between weak and strong 't Hooft coupling, by the introduction of
the so-called dressing phase
factor~\cite{Arutyunov:2004vx,Beisert:2005tm,Beisert:2006ez}.

Another question, that has received somewhat less attention, is to
what extent gauge theory and string theory agree to first order in
$\lambda'$. As stated above, agreement has been found up to first
order in $1/J$ \cite{Callan:2003xr}.%
\footnote{In \cite{Harmark:2008gm} an argument is given for why, at the order $\lambda'$, gauge
theory and string theory agree up to first order in $1/J$. This argument is based the so-called
decoupling limit \cite{Harmark:2006di,Harmark:2006ta,Harmark:2006ie,Harmark:2007px,Harmark:2007et} of
AdS/CFT.} However, the agreement has not been tested beyond first order in $1/J$. It has been
conjectured in \cite{Arutyunov:2004vx} that the planar gauge theory and tree-level string theory
agree exactly to all orders in $1/J$ in the sense that the same Bethe equations and dispersion
relation describe both. From the string theory point of view this is interesting since one should see
the emergence of the discrete nature of the string world-sheet from an $E=p^2$ to an $E=4\sin^2
(p/2)$ type of dispersion relation (in the $\mathfrak{su}(2)$ sector).

In this paper we explore this question for the case of the
semi-classical circular closed string state
\cite{Frolov:2003qc,Beisert:2003xu,Arutyunov:2003uj} and furthermore
for its orbifolded generalization. The circular string that we
consider is confined in a $\R \times S^3$ subspace of $\ads_5 \times
S^5$, with the $S^3$ being inside $S^5$. The circular string in this
subspace has two independent angular momenta $J_1$ and $J_2$
corresponding to the two rotation angles $\phi_1$ and $\phi_2$ of
the $S^3$. On the gauge theory side, these angular momenta are
identified with two of the $R$-charges. The string has large
$J=J_1+J_2$ but with the ratio $\alpha = J_2/J$ fixed. The circular
string is characterized by having a non-zero winding number $m$ with
respect to the angle $\varphi = \phi_1 - \phi_2$.

On the gauge theory side the circular string state is mapped to an
operator in the $\mathfrak{su}(2)$ sector, being $\tr ( Z^{J_1}
X^{J_2} )$ or permutations thereof, where $Z$ and $X$ are two of the
complex scalars of $\CN=4$ SYM theory. The one-loop scaling
dimensions of operators in the $\mathfrak{su}(2)$ sector are
described exactly by the $XXX_{1/2}$ ferromagnetic Heisenberg spin
chain and its corresponding Bethe equations.

One of the aims of this paper is to match higher order corrections
in $1/J$ between the circular string state and the corresponding
gauge theory operator, for the part of the energy which is
first-order in $\lambda'$. However, it is difficult to understand
such corrections in the full quantum string theory, since that
requires to include modes outside the $\R \times S^3$ subspace.
Instead, we adopt in this paper the approach of
\cite{Minahan:2005mx,Minahan:2005qj} and use the Landau-Lifshitz
sigma-model~\cite{Fradkin,Kruczenski:2003gt,Kruczenski:2004kw}, plus
certain higher derivative terms, as an effective description of the
string side. This is furthermore known to be a long wave-length
approximation to the $XXX_{1/2}$ ferromagnetic Heisenberg spin chain
which we have on the gauge theory side.

We consider first the circular string using the Bethe equations for
the $XXX_{1/2}$ ferromagnetic Heisenberg spin chain. This has
previously been considered
in~\cite{Beisert:2005mq,Hernandez:2005nf}. Since we are interested
in finding higher order corrections in $1/J$, we employ a novel way
of solving the Bethe equations. This consists of making an expansion
in powers of the winding number $m$ while at the same time having
$J$ large. We obtain in this way the $m^2$ and $m^4$ contributions
to the energy
\begin{equation}
\label{su2en} \frac{E - J}{\lambda'} = \frac{1-\alpha}{\alpha}
\frac{J m^2}{2(1-1/J)} \left(1-\frac{(1-\alpha)}{\alpha}\frac{\pi^2
m^2}{3(J-1)}\right)+\mathcal{O}(m^6)
\end{equation}
This result is consistent with previous results for circular strings
\cite{Beisert:2005mq,Hernandez:2005nf}. Notice that eq.~(\ref{su2en}) contains the corrections in
$1/J$ to any order. The novel procedure that we use to solve the Bethe equations takes advantage of
some exact properties of the zeroes of the Laguerre polynomial~\cite{Ahmed:1978uw} and, being quite
powerful, it could presumably be extended also to higher powers of the winding number of the string
states. However, it is important to remark that for finite $J$ one needs to add additional
contributions that are non-perturbative in $1/J$. These non-perturbative contributions are related to
the instabilities of the circular string \cite{Beisert:2005mq,Hernandez:2005nf,Minahan:2005mx}.

Since the solution of the Bethe equations for the $XXX_{1/2}$
Heisenberg spin chain, describing the one-loop contribution to the
$\mathfrak{su}(2)$ sector of $\CN=4$ SYM theory, is closely related
to the Bethe equations for the $XXX_{-1/2}$ Heisenberg spin chain,
which instead describes the one-loop contribution to the
$\mathfrak{sl}(2)$ sector, we solve both sectors at the same time.
The $\mathfrak{sl}(2)$ sector consists of operators of the type $\tr
( D^{s_1} Z D^{s_2} Z \cdots D^{s_J} Z )$. Here $D$ is $D_1+iD_2$,
$D_\mu$ being the covariant derivative, $S = s_1 + s_2 + \cdots +
s_J$ and $J$ is the number of $Z$'s. The string solution is in this
case called the spinning closed string since the string is spinning
in the $\ads_5$ space \cite{Arutyunov:2003uj,Arutyunov:2003za}. We
find for large $J$ and finite $\alpha=S/J$
\begin{equation}
\label{sl2en} \frac{E - S - J}{\lambda'} = \frac{1+\alpha}{\alpha}
\frac{J m^2}{2(1+1/J)} \left(1-\frac{(1+\alpha)}{\alpha}\frac{\pi^2
m^2}{3(J+1)}\right)+\mathcal{O}(m^6)
\end{equation}
This is consistent with previous results for spinning strings \cite{Beisert:2005mq}\footnote{The
problem of computing higher order finite size corrections has also been addressed in
~\cite{Gromov:2005gp,Gromov:2007ky}. It would be interesting to compare their results with ours.}.

The {\sl orbifolded} circular string solution, that we also consider
in this paper, is a generalization of the circular string solution
to string theory on $\ads_5\times S^5 / \Z_M$ so that the subspace
in which we have the string is $\R \times S^3 / \Z_M$. The dual
gauge theory is a $\mathcal{N}=2$ quiver gauge theory (QGT) with the
orbifolded circular string corresponding to a completely symmetrized
trace of $J_1$ complex scalars $Z$ and $J_2$ complex scalars $X$
with $J_1=J_2=J/2$ and a suitably inserted twist
matrix~\cite{Bertolini:2002nr}. The Bethe ansatz that provides the
anomalous dimension for these
operators~\cite{Ideguchi:2004wm,Beisert:2005he,Astolfi:2006is}
contains a twist depending on the winding through the ratio $m/M$
and its solution gives back the energy of the circular rotating
string by setting $M=1$. These operators cannot be directly
inherited from the parent $\mathcal{N} = 4$ theory, in fact, because
of the appearance of the twist matrix, the winding state involves
the twisted sectors of the $\mathcal{N}=2$ QGT. We generalize the
solution of the Bethe equations for the $\alpha=1/2$ circular string
to the orbifolded case. This is readily achieved and the result is
again (\ref{su2en}) with the substitution of $m$ with $m/M$.%
\footnote{In \cite{Larsen:2007bm} the $\mathfrak{su}(2)$ decoupling limit
of \cite{Harmark:2006di,Harmark:2006ta,Harmark:2006ie,Harmark:2007px,Harmark:2007et}
is generalized to orbifolded $\mathcal{N}=2$ quiver gauge theory.}

The {\sl winding state} is given by the limit of orbifolded circular
string solution with $M^2/J$ fixed for $J \rightarrow \infty$
\cite{Bertolini:2002nr}. In \cite{Bertolini:2002nr} a Penrose limit
of $\ads_5\times S^5 / \Z_M$ giving a pp-wave background with a
compact spatial direction (with 24 supersymmetries) is considered.
The winding state corresponds to a string winding around the compact
spatial direction.%
\footnote{The identification of the winding state studied in this paper and the winding
state of \cite{Bertolini:2002nr} is provided in detail in \cite{windingstring}.}
One of the reasons why it is interesting to study
this type of state is that, in the large $M$ limit, the
instabilities of the circular string are absent. Moreover since the
winding state on the string side is a vacuum state for the string
excitations, the energy found from the Bethe equations should be
reproduced purely by quantum string effects on the string theory
side.

In this paper we shall not only be concerned with solutions of the
Bethe equations but we will also try to match the results obtained
from the Bethe equations with those coming form the corresponding
coherent state ``Landau-Lifshitz" (LL) sigma  model which describes
low energy states of the ferromagnetic
spin-chain~\cite{Fradkin,Kruczenski:2003gt,Kruczenski:2004kw}. This
LL type action creates a connection between the gauge theory and the
string theory pictures, suggesting how a continuous string action
may appear from the gauge theory, as well as providing further
evidence of the microscopic spin-chain description of string theory.
The LL action is an effective low-energy action that arises from the
gauge theory - spin chain and the quantum superstring, and, as such,
it cannot be expected to lead to a well-defined quantum theory.
However, supplemented with an appropriate regularization
prescription and with higher-derivative counterterms, the LL model
has been able to capture a non-trivial part of the quantum
corrections to the ``microscopic'' theories, the string and
spin-chain~\cite{Minahan:2005mx,Minahan:2005qj}.

We will compare up to the order $1/J^2$ the energy for the circular
string and the winding state obtained from the Bethe equations with
those derived from the LL model. For the first two leading terms in
the winding number $m$, where the Bethe equations have been solved
at any order in $J$, using, in the case of the winding state, an
orbifolded version of the LL sigma-model, we will show that the
results of the two computations actually match.

For the circular string the first order correction in $1/J$ can be reproduced by $\zeta$-function
regularization, as found in \cite{Beisert:2005mq,Hernandez:2005nf,Minahan:2005mx}. This matches the
$m^2$ and $m^4$ corrections at first order in $1/J$ in (\ref{su2en}). We go on to compute, in two
different parametrizations of the LL model, the second order correction in $1/J$, again using
$\zeta$-function regularization. The two parametrizations, which give rise to rather different
effective Lagrangians, yield in a non-trivial fashion the same result (this happens only thanks to a
non-trivial
cancellation of divergences).%
\footnote{The result differs from \cite{Minahan:2005mx}, see Section \ref{circular}.} The result
matches the $m^2$ part of (\ref{su2en}), at second order in $1/J$, but, however, not the $m^4$ part.
This can be explained by the fact that the regularization that actually corresponds to the UV finite
microscopic theories not necessarily is the $\zeta$-function regularization, as suggested also
in~\cite{Minahan:2005mx,Minahan:2005qj}. Clearly, the most satisfactory way to resolve this question
would be to make a complete superstring calculation to this order, since the superstring sigma-model
should automatically choose the right regularization prescription.
% way to cancel the infinities coming the modes of the LL model.

In the case of the winding state we consider the orbifolded LL model
that arises either by taking a limit of the classical sigma-model on
$\R \times S^3 / \Z_M$, or from orbifolding the Bethe equations for
the $\mathfrak{su}(2)$ sector. While the target space of the LL
model is $S^2$ with rotation angle $\varphi = \phi_1 - \phi_2$, the
orbifolded LL model corresponds to the same sigma-model action but
with the identification $\varphi \equiv \varphi + 4\pi /M$. Taking
the limit of large $J$, with $M^2/J$ fixed, it reveals to leading
order a cylinder $S^1 \times \R$ as the target space for the
sigma-model. The winding state describes a string winding around the
compact direction.

We are able to match the leading order, the $1/J$ correction and the
$1/J^2$ correction to the energy of the winding state, as computed
from the Bethe equations and the orbifolded LL model. To first order
in $1/J$ we have a $m^2/J$ term only. This is matched by observing
that certain non-normal ordered terms in the $1/J$ Hamiltonian can
contribute to the energy due to the absence of a zero mode for the
string.

At $1/J^2$  we have to use second order perturbation theory. This
gives rise to two terms, an $m^2/J^2$ and an $m^4/J^2$ term. The
matching of both these two terms is highly interesting. The
$m^2/J^2$ term arises from the mean value of the interaction
Hamiltonian on the winding state. It can be matched by carefully
considering the ordering of the two coordinates that we use to
parameterize the target space. We show that irrespective of what
ordering convention we use, we always get the same answer for the
$m^2/J^2$ term. In particular, one can use Weyl
ordering~\cite{Szabo:2001kg}. The $m^4/J^2$ term comes instead by
summing over the set of intermediate states. We find that the only
non-zero contribution to the energy is found by summing over the
individual contribution of all the possible two-oscillator string
states created from the vacuum. We believe that this matching is
rather novel and non-trivial in this respect, in that it is the
first time that a successful match of gauge theory and a continuous
sigma-model has relied on summing over intermediate states with a
different number of oscillators compared to the external state.

This paper is organized as follows. In Section \ref{sec:finitespinning} we consider finite-size
corrections to the circular string state in the $\mathfrak{su}(2)$ sector and the spinning string in
the $\mathfrak{sl}(2)$ sector. In Section \ref{sec:bethesolution} we find the finite-size corrections
(\ref{su2en}) and (\ref{sl2en}) from the Bethe equations. We subsequently consider the finite-size
corrections in the $\mathfrak{su}(2)$ sector as computed from the LL model in Section \ref{circular}.
In Section \ref{sec:finitewinding} we consider the finite-size size corrections to the winding state,
first from the Bethe equations in Section \ref{sec:winbet} and subsequently from the orbifolded LL
model in Section \ref{sec:winLL}. We conclude and discuss future perspectives in Section
\ref{sec:concl}.

%%%%%%%%%%%%%%%%%%%%%%%%%%%%%%%%%%%%%%%%%%%%%%%%%%%%%%%%%%%%%
\sectiono{Finite-size corrections to the circular/spinning string state}
\label{sec:finitespinning}

Our aim here is to compute the quantum finite size corrections to
the one-loop anomalous dimensions of operators of the form
${\rm{Tr}}(Z^{J_1}X^{J_2})$ and ${\rm{Tr}}\left(D^{S} Z^J\right)$ in
the $\mathfrak{su}(2)$ and $\mathfrak{sl}(2)$ sectors, respectively.
These are conjectured to be equal to the energy of the circular
rotating string with two independent angular momenta $J_1$ and $J_2$
in  $S^5$, winding number $m$, and to the energy of the spinning
string with spin $S$ in $AdS_5$ and angular momentum $J$ on $S^5$.
The latter solution may be viewed as an analytic continuation of the
first so that the two cases can be treated simultaneously just by
keeping track of the sign differences in the Bethe equations for the
two sectors. Both the $\mathfrak{su}(2)$ sector and the
$\mathfrak{sl}(2)$ sectors are described by a $XXX$ Heisenberg spin
chain, the $\mathfrak{su}(2)$ with spin $1/2$ and the
$\mathfrak{sl}(2)$ with spin $-1/2$.

The Bethe equations will be solved by reformulating the problem in terms of
the resolvent function $G(x)$ as in~\cite{Lubcke:2004dg,Beisert:2005mq} but
by keeping into account also the so-called anomalous contribution arising
from the fraction of the Bethe roots whose distance is of order
$1/J$~\cite{Beisert:2005mq,Bargheer:2008kj}. We shall explore a different region
of the parameters compared to the one studied in~\cite{Beisert:2005mq}.
With $J$ large and $\alpha={K}/{J}$ fixed but finite, (here $K$ is the number
of impurities $i.e.$ $J_2$ for the $\mathfrak{su}(2)$ sector and $S$ for the
$\mathfrak{sl}(2)$ sector) we shall be able to go beyond the $1/J$ order result
of~\cite{Beisert:2005mq} and determine the spectrum for any value of $J$  in an
expansion in the winding number $m$.
This will be done by taking advantage of an exact property of the
zeroes of the Laguerre polynomial found in~\cite{Ahmed:1978uw}.

%%%%%%%%%%%%%%%%%%%%%%%%%%%%%%%%%%%%%%%%%%%%%%%%%%%%%%%%%%%%%
\subsection{All-order finite-size effects from Bethe equations}
\label{sec:bethesolution}

The spectrum of anomalous dimension of operators in the
$\mathfrak{sl}(2)$ and $\mathfrak{su}(2)$ sectors is given, at
one-loop, by the solution of the Bethe
equations~\cite{Minahan:2002ve}
\begin{equation}
\label{betheeqs} \left( \frac{u_k+\frac{i}{2} r }{u_k-\frac{i}{2} r}
\right)^J = \prod_{j\neq k} \frac{u_k - u_j + i}{u_k - u_j - i}
\end{equation}
where $r=-1$ for $\mathfrak{sl}(2)$ and $r=1$ for
$\mathfrak{su}(2)$. The indices $j$ and $k$ go from $1$ to $K$, $K$
being the magnon number. The Bethe equations describe completely the
spectrum of the spin chain and for $r=1$ are those of the Heisenberg
magnet.

If we take the logarithm of (\ref{betheeqs}), we get%
\footnote{Here and in the following we are considering the branch of
$\arctan$ with $\arctan(0)=0$.}
\begin{equation}
\label{gbeteq} \pi n + r J \arctan \left( \frac{1}{2u_k} \right) =
\sum_{j\neq k} \arctan \left( \frac{1}{u_k-u_j} \right)
\end{equation}
where $n \in \mathds{Z}$ reflects the arbitrariness in choosing the
branch of the logarithm. In general one can choose a different $n$
for each $k$, but we restrict ourselves here to the special case
where $n$ is the same for all $k$. The constraint from the cyclicity
of the trace gives the momentum condition
\begin{equation}
\label{defmm} m \equiv \frac{1}{\pi} \sum_{k=1}^K \arctan \left(
\frac{1}{2u_k} \right) \in \Z
\end{equation}
The one-loop contribution to the energy is
\begin{equation}
\label{oneloopen} \CE = \frac{\lambda}{8\pi^2} \sum_k \frac{1}{u_k^2
+ \frac{1}{4}}
\end{equation}
where $\CE = E-J$ for the $\mathfrak{su}(2)$ sector and $\CE=E-S-J$ for the $\mathfrak{sl}(2)$
sector, with $E$ being the full scaling dimension of the operator. In the $\mathfrak{su}(2)$ sector
$J=J_1+J_2$ where $J_1$ and $J_2$ are two of the R-charges. In the $\mathfrak{sl}(2)$ sector $J$ is
an R-charge while $S$ is an angular momentum. Summing over all $k$ in (\ref{gbeteq}) gives zero on
the right-hand side. Therefore, we get the constraint
\begin{equation}
\label{ncon} K n + r J m = 0
\end{equation}
This is as in string theory, it provides the level matching
condition in the presence of a winding. We define $\alpha$ as
\begin{equation}
\alpha \equiv \frac{K}{J}
\end{equation}
Therefore $n = - r m /\alpha$. The string requires both $n$ and $m$
to be integers so that $\alpha$ can only be a divisor of $m$, for
the spin-chain we can instead consider states with any value of
$\alpha$.

It is apparent that the one-loop contribution to the energy (\ref{oneloopen}) of the state defined by
the Bethe equations (\ref{gbeteq}) and the momentum constraint (\ref{defmm}) has the functional form
$\CE= \lambda f(J,\alpha,m)$. The full function $f(J,\alpha,m)$ is unknown. In previous works on the
subject~\cite{Lubcke:2004dg,Beisert:2005mq}, $f(J,\alpha,m)$ has been expanded as
\begin{equation}
\label{Jexpansion} f(J,\alpha,m) = \frac{1}{J} f_1(\alpha,m) +
\frac{1}{J^2} f_2(\alpha,m) + \frac{1}{J^3} f_3(\alpha,m) + \cdots
\end{equation}
In this approach $f_1(\alpha,m)$ and $f_2(\alpha,m)$ have been
found, revealing the following expression for the one-loop
contribution to the energy
\begin{equation}
\label{prevresult} \CE = \frac{\lambda'\,J \hat{m}^2}{2} +
r\frac{\lambda' \hat{m}^2}{2} + \frac{\lambda'}{2} \sum_{n=1}^\infty
\left[ n^2 \sqrt{1-r\frac{4\hat{m}^2}{n^2} } -n^2 + 2r \hat{m}^2
\right] + \CO( \lambda' J^{-1} )
\end{equation}
where we defined
\begin{equation}
\label{hatm} \hat{m} \equiv \sqrt{ \frac{1-r\alpha}{\alpha} } m
\end{equation}

In the following we propose instead to consider the expansion
\begin{equation}
\label{mexpansion} f(J,\alpha,m) = m^2 g_1(J,\alpha) + m^4
g_2(J,\alpha) + m^6 g_3(J,\alpha) + \cdots
\end{equation}
Using this way of expanding $f(J,\alpha,m)$ we will be able to
determine $g_1(J,\alpha)$ and $g_2(J,\alpha)$. However, as we shall
see below, the result is only reliable for $K = \alpha J \gg m$
since otherwise one cannot make sense of this expansion.

Obviously (\ref{mexpansion}) is only an expansion in a formal sense since if for example one sets
$m=1$ then one needs all the infinite number of terms in the expansion (\ref{mexpansion}). The
expansion is nevertheless still useful since one can keep $m$ free. Moreover, in Section
\ref{sec:winbet} we shall show that in the orbifolded theory only a finite number of terms in the
expansion (\ref{mexpansion}) contributes to a given power of $1/J$ in a $1/J$ expansion, thus making
it sensible also to set $m=1$.

Define now
\begin{equation}
x_k \equiv \frac{2\pi r n}{J} u_k = - \frac{2\pi m}{\alpha J} u_k
\end{equation}
and the expansion parameter $\epsilon$
\begin{equation}\label{epsilon}
\epsilon \equiv \frac{4\pi^2 m^2}{\alpha^2 J^2}=\frac{4\pi^2
m^2}{K^2}
\end{equation}
which is small,  $\epsilon \ll 1$, only for $K = \alpha J \gg m$,
$i.e.$ we need a large magnon number.

Then (\ref{gbeteq}) can be expanded in $\epsilon$ as
\begin{equation}
\label{Nxkeqs} 1 + \frac{1}{x_k} -  \frac{\epsilon}{12x_k^3} +
\frac{\epsilon^2}{80x_k^5} + \cdots  = \frac{r}{J} \sum_{j \neq k}
\left( \frac{2}{x_k-x_j} -  \frac{2\epsilon}{3 (x_k-x_j)^3 } +
\frac{2\epsilon^2}{5(x_k-x_j)^5} +\cdots \right)
\end{equation}
We furthermore have that the momentum constraint (\ref{defmm}) is expanded as
\begin{equation}
\sum_k \left( \frac{1}{x_k} - \epsilon \frac{1}{12x_k^3} + \cdots
\right) = - \alpha J
\end{equation}

Define the resolvent
\begin{equation}
G(x) = \frac{1}{J} \sum_{k=1}^K \frac{1}{x-x_k}
\end{equation}
We expand $G(x)$ in powers of $\epsilon$ as
\begin{equation}
\label{NGep} G(x) = G_0(x) + \epsilon G_1(x) + \epsilon^2 G_2(x) +
\cdots
\end{equation}
The momentum constraint to first order in $\epsilon$ gives
\begin{equation}
\label{momconn} G_0(0) = \alpha \spa G_1(0) = \frac{1}{24} G''_0(0)
\end{equation}
To first order in $\epsilon$ we can write the Bethe equations (\ref{gbeteq}) as
\begin{eqnarray}
\label{newgeqs} r G(x)^2 + \frac{r}{J} G'(x)  &= & - \frac{\alpha}{x} + \left( 1 + \frac{1}{x} -
\frac{\epsilon}{12x^3} \right) G(x) + \frac{\epsilon}{12} \left( \frac{\alpha}{x^3} +
\frac{G'(0)}{x^2} \right) \cr & + & r \frac{2\epsilon}{3J^2} \sum_k \frac{1}{x-x_k} \sum_{j \neq k}
\frac{1}{(x_k-x_j)^3 } + \CO(\epsilon^2)
\end{eqnarray}
Using (\ref{NGep}) we see that $G_0(x)$ should obey the equation
\begin{equation}
\label{newg0eq} r G_0(x)^2 + \frac{r}{J} G_0'(x) = -
\frac{\alpha}{x} + \left( 1 + \frac{1}{x} \right) G_0(x)
\end{equation}
Introducing the function
\begin{equation}
\label{Q} Q(x)=\prod_{i=1}^K(x-x_i)
\end{equation}
known as the eigenvalue of the Baxter $Q$-operator, we can write now
\begin{equation}
G(x) = \frac{1}{J} \frac{Q'(x)}{Q(x)}
\end{equation}
Then, using (\ref{NGep}) at the lowest order in $\epsilon$, (\ref{newg0eq}) is equivalent to
\begin{equation}
\label{diffQ} \left[ x \frac{d^2}{dx^2} - r J (x+1) \frac{d}{dx} + r
\alpha J^2 \right] Q = 0
\end{equation}
It is useful to define the variable $y \equiv rJ x$. We see that (\ref{diffQ}) written in terms of
$y$ is equivalent to the Laguerre differential equation (\ref{lagpol}) given in Appendix
\ref{app:laguerre} with $\nu = - rJ - 1$ and $\lambda = \alpha J$. This means that we have the
solution
\begin{equation}
Q(x) \propto L^{-rJ-1}_{\alpha J} ( r J x )
\end{equation}
where $L^\nu_\lambda(y)$ is the Laguerre polynomial, see Appendix \ref{app:laguerre}. We can now use
the sum rules (\ref{zerosum1}) and (\ref{zerosum3}) for the zeroes of Laguerre polynomials
\cite{Ahmed:1978uw}. We label the zeroes as $y_k$, $k=1,2,..., K$. It is not difficult to see that
(\ref{zerosum1}) is equivalent to the zeroth order part of (\ref{Nxkeqs}), $i.e.$
\begin{equation}
\frac{2r}{J} \sum_{j \neq k} \frac{1}{x_k-x_j} = 1 + \frac{1}{x_k}
\end{equation}
when setting $y_k = r J x_k$. Thus we have a clear connection between the Bethe roots $x_k$ and the
zeroes of an associated Laguerre polynomial. Instead the sum (\ref{zerosum3}) gives
\begin{equation}
\sum_{j \neq k} \frac{1}{(x_k-x_j)^3} = - \frac{J(J-2r)}{8x_k^3} -
\frac{J^2(1-2\alpha r)}{8x_k^2}
\end{equation}
Using this in (\ref{newgeqs}) one obtains the following equation for $G_1(x)$
\begin{eqnarray}
\label{newg1eq} && \big[ 2 r x G_0(x) - x - 1 \big] G_1(x) + \frac{rx}{J}G_1'(x) \cr && = -
\frac{1}{12} \left( \frac{rJ-1}{x^2} + \frac{rJ(1-2\alpha r)}{x} \right) \left[ G_0(x) - \alpha -
xG_0'(0) \right] + \frac{r J - 2}{24} G''_0(0)
\end{eqnarray}

The energy $\CE$ is computed using
\begin{equation}
\label{Genergy}   \sum_k \frac{1}{u_k^2+\frac{1}{4}} = - \epsilon J
G_0'(0) + \epsilon^2 J \left[ - G_1'(0) + \frac{1}{24} G'''_0(0)
\right] + \CO(\epsilon^3)
\end{equation}
We see that all we need to know in order to find the energy is
$G_0'(0)$, $G'''_0(0)$ and $G_1'(0)$. $G_0'(0)$ and $G'''_0(0)$ are
found by writing $G_0(x)$ in a Taylor expansion
\begin{equation}
\label{taylorG0} G_0(x) = \frac{1}{2} + G_0'(0) x + \frac{1}{2}
G_0''(0) x^2 + \frac{1}{6} G_0'''(0) x^3 + \cdots
\end{equation}
Inserting this into (\ref{newg0eq}) and expanding in $x$ we find
\begin{equation}
G_0'(0) = - \frac{(1-r \alpha) \alpha  J}{J-r} \spa G_0''(0) =
\frac{2\alpha J^2 (1-r \alpha)(1-2r \alpha)}{(J-r)(J-2r)}
\end{equation}
\begin{equation}
G_0'''(0) = - \frac{6\alpha J^3 (1-r\alpha)
[J(1-5r\alpha+5\alpha^2)-r+6\alpha-6r\alpha^2]}{(J-r)^2
(J-2r)(J-3r)}
\end{equation}
Making a similar Taylor expansion for $G_1(x)$ as in (\ref{taylorG0}) we get from (\ref{newg1eq})
\begin{equation}
G_1'(0) = \frac{Jr}{72} [ G_0'''(0) + (3-6 r \alpha) G_0''(0) ]
\end{equation}
Using this in (\ref{Genergy}) we get
\begin{equation}
\label{finE} \CE =  \frac{\lambda'\,J^2 \hat{m}^2}{2(J-r)} -
\frac{\pi^2}{6} \frac{\lambda'\,J^2 \hat{m}^4}{(J-r)^2} + \CO( m^6)
\end{equation}
where $\hat{m}$ is defined in (\ref{hatm}). We see thus that we have obtained $g_1(J,\alpha)$ and
$g_2(J,\alpha)$ in the formal expansion (\ref{mexpansion}) in powers of $m$, as promised. It is
useful to recall here the validity of this equation.  We chose an expansion parameter $\epsilon$
defined in (\ref{epsilon}), which, in order to be small, requires a large number of impurities $K \gg
m$. At the same time we expanded in the variable $m=-r \alpha n$ so that $\alpha$ has to be kept
finite, it cannot be sent to infinity for example as in the case studied in~\cite{Beisert:2005mq}.
However the coefficient of the $m^2$ and $m^4$ terms $g_1(J,\alpha)$ and $g_2(J,\alpha)$ for fixed
finite $\alpha$ provide the finite size corrections to all orders in $J$. In the next section, using
the Landau-Lifshitz model, we will be able to match, using this formula, the $\hat m^2/J^2$ term
which could not be derived in~\cite{Beisert:2005mq}. Expanding (\ref{finE}) we get
\begin{equation}\label{finEexp}
\CE= \lambda' J \left(\frac{\hat{m}^2}{2} + r \frac{\hat{m}^2}{2J} +
\frac{\hat{m}^2}{2J^2} -
\frac{\pi^2}{6}\frac{\hat{m}^4}{J}-r\frac{\pi^2}{3}
\frac{\hat{m}^4}{J^2}\right) + \CO(m^6)    + \CO( \lambda' J^{-2} )
\end{equation}
We can now compare the result (\ref{finE}) with the result (\ref{prevresult}) for the one-loop energy
contribution obtained in \cite{Beisert:2005mq} in the $1/J$ expansion (\ref{Jexpansion}). Expanding
(\ref{prevresult}) in powers of $\hat m$, we obtain
\begin{equation}
\CE= \lambda' J\left( \frac{\hat{m}^2}{2} +r \frac{\hat{m}^2}{2J} -
\zeta(2) \frac{\hat{m}^4}{J}- 2 r \zeta(4)
\frac{\hat{m}^6}{J}\right) + \CO(m^8)    + \CO( \lambda' J^{-1} )
\label{finalbethe}
\end{equation}
which is seen to match (\ref{finEexp}), for the common terms.

%%%%%%%%%%%%%%%%%%%%%%%%%%%%%%%%%%%%%%%%%%%%%%%%%%%%%%%%%%%%%
\subsection{Results from Landau-Lifshitz sigma-model}
\label{circular}

A convenient framework where to compute finite size corrections to
the energy of a given state is provided by the Landau-Lifshitz (LL)
model~\cite{Fradkin,Kruczenski:2003gt,Kruczenski:2004kw}.

The LL model was introduced in the study of the low energy spectrum
of the ferromagnetic Heisenberg spin chain. The Sigma-model action
is given by
\begin{equation}
\label{II} I =  \frac{\lambda' J }{4\pi } \int d t \int_0^{2\pi} d\sigma \left[ \vec{C}(\vec{n})
\cdot \dot{\vec{n}} - \frac{1}{4} \left( \vec{n}' \right)^2  \right]
\end{equation}
where $\vec{n}$ is a three-dimensional unit-vector parameterizing the two-sphere, the ``prime'' means
derivative with respect to $\sigma$ and the  ``dot" derivative with respect to $t$. Note that the
first term in (\ref{II}) is a Wess-Zumino type term which is proportional to the area spanned between
the trajectory and the north pole of the two-sphere \cite{Fradkin}. Choosing the parametrization
\begin{equation}
\vec{n} = (\cos \theta \cos \varphi, \cos \theta \sin \varphi , \sin
\theta) \label{vet}
\end{equation}
we have that
\begin{equation}
\vec{C}(\vec{n}) \cdot \dot{\vec{n}} = \sin \theta \dot{\varphi} \spa  (\vec{n}')^2 = (\theta')^2 + \cos^2 \theta (\varphi')^2
\end{equation}
The action (\ref{II}) corresponds to the Hamiltonian
\begin{equation}
\label{ham1}
H = \frac{\lambda' J}{4\pi } \int_0^{2\pi} d\sigma \frac{1}{4} ( \vec{n}' )^2
\end{equation}

Classically, a circular rotating string is a configuration of the
type
\begin{equation}
\label{clascirco} \theta = 0 \spa \varphi = 2m\sigma
\end{equation}
where $m$ is an integer winding number. We want to study
fluctuations around this solution.
This LL solution corresponds to the term of order $\lambda$ in the
circular string solution \cite{Frolov:2003qc,Arutyunov:2003za}. The
classical energy, obtained by expanding the energy of the full
solution, $E=\sqrt{J^2+\lambda m^2}$, is given by
\begin{equation}
E_0=J\left(1+\lambda' \frac{m^2}{2}+\mathcal{O}({\lambda'}^2)\right)
\label{clascircen}
\end{equation}

The sigma-model action as derived from the Heisenberg spin chain also contains higher derivative
terms coming in at higher powers in $1/J$. Adding the first higher-derivative term, the Hamiltonian
(\ref{ham1}) changes to
\begin{equation}
\label{ham2}
H = \frac{\lambda' J}{4\pi } \int_0^{2\pi} d\sigma \left[
\frac{1}{4} ( \vec{n}' )^2 - \frac{\pi^2}{12 J^2} (\vec{n}'')^2 +
\CO (J^{-4}) \right]
\end{equation}
We record that
\begin{eqnarray}
( \vec{n}'' )^2 &=&  (\theta'')^2 + \cos^2 \theta (\varphi'')^2 +
(\theta')^4 + \cos^2 \theta (\varphi')^4 + 2 (2- \cos^2 \theta)
(\theta')^2 (\varphi')^2 \nn \\ &&
 + 2\sin \theta \cos \theta [ \theta'' (\varphi')^2 - 2 \theta' \varphi'
 \varphi'' ]
\end{eqnarray}
Evaluated on the  classical state (\ref{clascirco}) the Hamiltonian (\ref{ham2}) gives
\begin{equation}\label{classicphd}
H = \frac{\lambda' J}{4\pi} \int_0^{2\pi} d\sigma \left[ \frac{1}{4}
( \varphi' )^2 - \frac{\pi^2}{12 J^2} (\varphi')^4 + \CO (J^{-4})
\right] = \frac{\lambda' J m^2}{2} - \frac{2\pi^2\lambda'  m^4}{3 J}
+ \CO (\lambda' J^{-3})
\end{equation}
The first term reproduces the classical energy at this order in $\lambda'$, (\ref{clascircen}), the
second term has to be viewed as a quantum counterterm added in order to match the discrete spin chain
result. We see that the higher derivative terms start contributing at the order $\lambda'/J$.

To study the fluctuations we denote the ground state for this LL model as $|0_m\rangle$. We want to
compute quantum corrections to the ground state energy. To do this, we expand the LL action
(\ref{II}) around the solution (\ref{clascirco}) and then we quantize the Hamiltonian for the
fluctuations. We write this as
\begin{equation}
\label{param}
\varphi = 2m \sigma +\frac{2 f}{\sqrt{J}}
\spa
\sin \theta = \frac{2g}{\sqrt{J}}
\end{equation}
Here we parameterized the fluctuations using the functions $f(t,\sigma)$ and $g(t,\sigma)$
corresponding to the two spherical angles on the two-sphere. In Appendix \ref{appB} we consider
another parametrization of the two-sphere leading to equivalent results. The parametrization
(\ref{param}) gives the Lagrangian
\begin{equation}
\label{fgcirclag} J\CL = 4 g \dot{f} - \CH_2 -\frac{1}{\sqrt{J}}
\CH_3 - \frac{1}{J} \CH_4 + \cdots
\end{equation}
with
\begin{equation}\label{calh2}
\CH_2 = ( f')^2 + (g')^2 -4m^2 g^2
\end{equation}
\begin{equation}
\CH_3 = - 8 m f'g^2\label{h3circo}
\end{equation}
\begin{equation}
\CH_4 = 4 g^2(g'^2-f'^2) \label{h4circo}
\end{equation}
The Hamiltonian is given by
\begin{equation}
H = \frac{\lambda'}{4\pi } \int_0^{2\pi} d\sigma \left( \CH_2
+\frac{1}{\sqrt{J}} \CH_3 + \frac{1}{J} \CH_4 + \cdots  \right)
\end{equation}
The equations of motion are \cite{Minahan:2005mx}
\begin{equation}
\dot f=-\frac{1}{2}\left(g''+4 m^2 g\right), ~~~~~~\dot
g=\frac{1}{2}f'' \label{circeom}
\end{equation}
The solution can be written as
\begin{eqnarray}
f(\tau,\sigma)&=&\frac{1}{2}\sum_{n=-\infty\atop n\ne
0}^{\infty}\sqrt{w_n}(a_n
e^{-i\omega_n\tau+in\sigma}+a_n^{\dagger}e^{i\omega_n\tau-in\sigma})\cr
g(\tau,\sigma)&=&-\frac{i}{2}\sum_{n=-\infty \atop n\ne
0}^{\infty}\frac{1}{\sqrt{w_n}}(a_n
e^{-i\omega_n\tau+in\sigma}-a_n^{\dagger}e^{i\omega_n\tau-in\sigma})
\label{fgcirc}
\end{eqnarray}
where
\begin{equation}
\omega_n=\frac{1}{2}
n\sqrt{n^2-4m^2},~~~~~~w_n=\sqrt{1-\frac{4m^2}{n^2}},~~~~~n=\pm
1,\pm 2,\ldots \label{freq}
\end{equation}
We impose the commutation relations
\begin{equation}
\left[f(\tau,\sigma),f(\tau,\sigma')\right]=0,~~~~~\left[g(\tau,\sigma),g(\tau,\sigma')\right]=0,~~~~~
\left[f(\tau,\sigma),g(\tau,\sigma')\right]=i\pi
\delta(\sigma-\sigma')-\frac{i}{2} \label{comm}
\end{equation}
where the zero mode contribution has been subtracted. These give
\begin{equation}
[a_n,a_k^{\dagger}]=\delta_{n-k}
\end{equation}
Note that with this choice of coordinates that parametrize the LL Lagrangian, even if $f$ and $g$ do
not commute, (\ref{comm}), there are no ordering problems in the definitions of (\ref{h3circo}) and
(\ref{h4circo}), $g$ in fact commutes with $f'$.

As it is well-known \cite{Frolov:2003qc, Arutyunov:2003za, Minahan:2005mx}, the solution
(\ref{fgcirc}) has unstable fluctuation modes with $n=\pm 1,\ldots,\pm 2m$. They are a manifestation
of an instability of the full homogeneous string solution. We will ignore these instabilities. Let us
just comment that in the case of the winding state which will be considered in the next section and
which is obtained by replacing $m$ with $\frac{m}{M}$, there are no unstable modes or instabilities.
More details can be found in the next section.

Another way to avoid instabilities is to consider, instead of the
$\mathfrak{su}(2)$ sector discussed in this section, the stable
solution of the $\mathfrak{sl}(2)$ sector, where one of the angular
momenta is on $AdS_5$ and the other one is on $S^5$
\cite{Arutyunov:2003za,Stefanski:2004cw,Bellucci:2004qr,SchaferNameki:2005tn}.

\subsection*{Leading correction to the circular string solution}

Using the solutions (\ref{fgcirc}), the quadratic Hamiltonian $H_2$, which gives the leading
correction to the energy of the ground state $|0_m\rangle$, can be written as
\begin{equation}
H_2=\frac{\lambda'}{4\pi}\int_0^{2\pi}d\sigma
\CH_2=\frac{\lambda'}{2}\sum_{n=-\infty}^{\infty}|\omega_n|(a_n
a_n^{\dagger}+a_n^{\dagger}a_n) \label{H2circ}
\end{equation}
and we then get
\begin{equation}
E_1= \langle
0_m|H_2|0_m\rangle=\frac{\lambda'}{2}\sum_{n=-\infty}^{\infty}|\omega_n|
\end{equation}
We see that the sum in the previous expression is divergent and
needs to be regularized. A natural regularization choice is to
subtract and add the divergent contribution and then, for the
latter, to use the $\zeta$-function regularization. We
obtain~\cite{Minahan:2005mx}
\begin{equation}
E_1=\frac{\lambda'}{2}\left[m^2+\sum_{n=1}^{\infty}(n\sqrt{n^2-4m^2}-n^2+2m^2)\right]
\label{E1circ}
\end{equation}
This result is in agreement with the full string theory 1-loop computation
\cite{Frolov:2003tu,Frolov:2004bh, Park:2005ji} and with the Bethe ansatz computation on the spin
chain side \cite{Beisert:2005mq,Beisert:2005bv}. We see that (\ref{E1circ}) expanded in powers of
$m$, up to and including the $m^4$ term, agrees with (\ref{finE}) for the leading terms in a large
$J$ expansion.

\subsection*{Next to leading correction to the circular string
solution}

Now we compute the next subleading correction to the energy of the
circular string. This corresponds to the $\lambda'/J$ result
obtained from the Bethe ansatz and to the 2-loop correction computed
on the string theory side. We use standard perturbation theory.
Since $\langle 0_m|H_3|0_m\rangle=0$, the first non trivial
correction to the energy of the ground state is given by the second
order perturbation theory formula
\begin{equation}
E_2=\sum_{|i\rangle \neq |0_m\rangle}\frac{\big|\langle
0_m|H_3|i\rangle \big|^2}{E_{0_m}-E_i}+\langle 0_m|H_4|0_m\rangle
=E_2^a+E_2^b \label{seccirc}
\end{equation}
where $|i\rangle$ is an intermediate state.

We start by computing $E_2^a$. We have that
\begin{equation}
H_3=-\frac{\lambda'}{\pi \sqrt{J}}m\int_0^{2\pi}d\sigma 2f'g^2
\end{equation}
We want to compute $\langle 0_m|H_3|i\rangle $, where the
intermediate state $|i\rangle$ is given by
\begin{equation}
|i\rangle=\frac{1}{\sqrt{\mathcal{N}}}\,a_{l}^{\dagger}a_{p}^{\dagger}a_{q}^{\dagger}|0_m\rangle
\label{intstate}
\end{equation}
where $\mathcal{N}$ is a normalization constant. We see that it is
necessary to chose intermediate states for which the number of
oscillators is different from the one of the external state. The
relevance of keeping into account the contributions coming from this
channel was stressed not only in Ref.\cite{Minahan:2005mx}, in a
similar LL type calculation, but also in the pp-wave string field
theory context in Refs.~\cite{Grignani:2005yv,Grignani:2006en}. We
get
\begin{eqnarray}
&&E_2^{a}=-\frac{1}{3!}\sum_{l\neq p\neq q}\frac{\big|\langle 0_m|H_3\,
a_{l}^{\dagger}a_{p}^{\dagger}a_{q}^{\dagger}|0_m\rangle\big|^2}{\lambda'
(|\omega_l|+|\omega_p|+|\omega_q|)} -\frac{1}{2}\sum_{l\neq p}\frac{\big|\langle 0_m|H_3\,
a_{l}^{\dagger}a_{p}^{\dagger}a_{p}^{\dagger}|0_m\rangle\big|^2}{\lambda' (|\omega_l|+2|\omega_p|)}
\cr &=& -\frac{m^2 \lambda'}{3 J}\sum_{l\neq p\neq
q}\delta(l+p+q)\frac{\left(l\sqrt{\frac{w_l}{w_pw_q}}+p\sqrt{\frac{w_p}{w_lw_q}}+q{\sqrt\frac{w_q}{w_lw_p}}\right)^2}
{l^2w_l+p^2w_p+q^2w_q}\cr&&-\frac{m^2 \lambda'}{2J}\sum_{l\neq
p}\delta(l+2p)\frac{\left(l\sqrt{\frac{w_l}{w_p^2}}+2p\sqrt{\frac{1}{w_l}}\right)^2}
{l^2w_l+2p^2w_p}\cr &&\label{E21circ}
\end{eqnarray}
where the sums go from $-\infty$ to $\infty$. Here we have separated the contribution obtained when
$l,p,q$ are all different, which is given by the first term on the r.h.s. of (\ref{E21circ}) both in
the first and the second line, and the contribution obtained when two integers among $l,p,q$ are
equal, which is given by the second term of (\ref{E21circ}).

We now proceed to compute $E_2^b$. We have that
\begin{eqnarray}
&&H_4=\frac{\lambda'}{4\pi J}\int_0^{2\pi}d\sigma \left[4g^2(g'^2-f'^2) \right]
=-\frac{\lambda'}{J}\sum_{l,p,q,s=-\infty}^{\infty} \frac{lp}{8\sqrt{w_lw_pw_qw_s}}\delta(l+p+q+s)\cr
&& \left[\left(1+w_lw_p\right)
(a_sa_qa_{-l}^{\dagger}a_{-p}^{\dagger}+a_{-s}^{\dagger}a_{-q}^{\dagger}a_la_p +a_sa_qa_{l}a_{p}
+a_{-s}^{\dagger}a_{-q}^{\dagger}a_{-l}^{\dagger}a_{-p}^{\dagger}\right.\cr
&-&\left.a_sa_{-q}^{\dagger}a_la_p-a_sa_{-q}^{\dagger}a_{-l}^{\dagger}a_{-p}^{\dagger}-
a_{-s}^{\dagger}a_qa_la_p-a_{-s}^{\dagger}a_qa_{-l}^{\dagger}a_{-p}^{\dagger})+ \left(1-w_lw_p\right)
\right.\cr &&\left.\left(
a_{-s}^{\dagger}a_qa_{-l}^{\dagger}a_p+a_sa_{-q}^{\dagger}a_{-l}^{\dagger}a_p
+a_{-s}^{\dagger}a_qa_la_{-p}^{\dagger}+ a_sa_{-q}^{\dagger}a_la_{-p}^{\dagger}-
a_sa_qa_{-p}^{\dagger}a_p-a_sa_qa_la_{-p}^{\dagger}\right )\right]\cr &&
\end{eqnarray}
We get~\footnote{The same computation has been done in~\cite{Minahan:2005mx}. However our result for
$\langle 0_m|H_4|0_m\rangle$ differs from the one of~\cite{Minahan:2005mx}. We show in
App.~\ref{appB} that, using a different parametrization, we obtain the same result for $\langle
0_m|H_4|0_m\rangle$ as obtained in this section. Moreover our result expanded in $m^2$ reproduces
exactly the  result (\ref{finE}) derived from the Bethe ansatz.}
\begin{equation}
\langle 0_m|H_4|0_m\rangle=\frac{\lambda'}{8
J}\sum_{l,p=-\infty}^{\infty}l^2\left(
\frac{1}{w_lw_p}-\frac{w_l}{w_p}\right)
\end{equation}
The divergent sums in the previous expression can be regularized as
before by adding and subtracting the divergent contribution and then
using $\zeta$-function regularization~\cite{Minahan:2005mx}. The
result is
\begin{equation}
E_2^b=\langle 0_m|H_4|0_m\rangle=\frac{\lambda'}{4J}\left[
\sum_{l=1}^{\infty}\left(\frac{l^2}{w_l}-l^2w_l -4m^2\right)
-2m^2\right] \left[ 2\sum_{p=1}^{\infty}\left(\frac{1}{w_p}-1\right)
-1\right] \label{e4circ}
\end{equation}

The total energy at this order is then given by the sum of (\ref{classicphd}), (\ref{E1circ}),
(\ref{E21circ}) and (\ref{e4circ}) and has the form~\cite{Minahan:2005mx}
\begin{equation}
E=J\left[1+\frac{m^2\lambda'}{2}\left(1+\frac{c_1}{J}+\frac{c_2}{J^2}+
\mathcal{O}\left(\frac{1}{J^3}\right)\right)+\mathcal{O}(\lambda'^2)\right]
\end{equation}
The coefficients $c_1$ and $c_2$ given by regularized sums may be evaluated numerically as
in~\cite{Minahan:2005mx}. Taking $m=1$ (and ignoring imaginary contributions of unstable modes) we
get~\footnote{The difference in the value of $c_2$ compared to the one obtained in
\cite{Minahan:2005mx} is due to the different form for the term (\ref{e4circ}) (see the previous
footnote), the normalization of the intermediate states (\ref{intstate}) and the inclusion of the
higher derivative term~(\ref{classicphd}).}
\begin{equation}
c_1=-0.893~,~~~~~~c_2=-5.44
\end{equation}

We now want to compare the result for $E_2$ with the Bethe ansatz result (\ref{finE}) . For this
case, being the angular momenta $J_1$ and $J_2$ equal, we have to set $\alpha=1/2$ and $r=1$ so that
$\hat m=m$. As before, we expand the energy in powers of $m^2$. We have
\begin{equation}
E_{2}=\frac{\lambda'}{J}\left(\frac{m^2}{2}-\frac{2}{3}\pi^2m^4\right)+\mathcal{O}(m^6)
+\CO( \lambda' J^{-2} )
\end{equation}
Putting together all the contributions, including the higher derivative term appearing in
(\ref{classicphd}) up to the order $m^4$, we obtain the following expression for the energy of the
circular string
\begin{equation}
E=J \left[ 1 +
\lambda'\left(\frac{m^2}{2}+\frac{m^2}{2J}+\frac{m^2}{2J^2}-\frac{\pi^2}{6}\frac{m^4}{J}
-\frac{4\pi^2}{3}\frac{m^4}{J^2}\right) \right]+\mathcal{O}(m^6)
+\CO( \lambda' J^{-2} ) \label{zerocirc}
\end{equation}
We see that this result coincides exactly with the one obtained from the Bethe ansatz, eq.
(\ref{finEexp}), for the terms proportional to $m^2$.

The coefficient of the $m^4/J$ term is also reproduced but from this
LL computation we find a mismatch for the coefficient of the term
proportional to $m^4/J^2$. This discrepancy is probably due to the
fact that the $\zeta$-function regularization is not appropriate at
the order $1/J^2$ (see also \cite{Minahan:2005mx} for comments on
this point).

Let us finally comment that in App. \ref{appB} the same computation is performed using a different
parametrization and it is shown that, using the same procedure described above, we get again the
result (\ref{zerocirc}). This result arises also thanks to a non-trivial cancelation of divergences.

%%%%%%%%%%%%%%%%%%%%%%%%%%%%%%%%%%%%%%%%%%%%%%%%%%%%%%%%%%%%%
\sectiono{Finite-size corrections to the winding state}
\label{sec:finitewinding}

In Section \ref{sec:finitespinning} we considered the finite-size
effects for the spinning/circular string state both from the Bethe
equations and from the Landau-Lifshitz model point of view. In this
section we generalize the circular string state to an orbifolded
circular string state. Geometrically, this means that whereas before
we were considering a string winding around an $S^3$ we consider
instead now a string winding around $S^3/\Z_M$. For $M=1$ it reduces
to the circular string state considered in Section
\ref{sec:finitespinning}. Instead when considering a limit of large
$M$ and large $J$, such that $M^2/J$ is finite, the orbifolded
circular string state becomes a winding state. This winding state is
related to a winding state for a particular pp-wave background
considered in \cite{Bertolini:2002nr} (see also
\cite{windingstring}).

In Section \ref{sec:winbet} we introduce the orbifolded circular
string state and we generalize the result of Section
\ref{sec:bethesolution} on the Bethe equations to this case. Then in
Section \ref{sec:winLL} we consider the orbifolded Landau-Lifshitz
model and we match the finite-size corrections to the energy found
from the Bethe equation to order $1/J^2$.

%%%%%%%%%%%%%%%%%%%%%%%%%%%%%%%%%%%%%%%%%%%%%%%%%%%%%%%%%%%%%
\subsection{Winding state from Bethe equations}
\label{sec:winbet}

In this section we generalize the result of Section
\ref{sec:bethesolution} on the Bethe equations to the case of the
orbifolded circular string state.

\subsubsection*{$\CN=2$ Orbifolded quiver gauge theory}

Consider the $\C^2 / \Z_M \times \C$ orbifold defined by the
identification
\begin{equation}
\label{orbifold} (Z,X,W) \equiv (\theta Z, \theta^{-1} X, W)
\end{equation}
where $Z$, $X$ and $W$ are three complex scalars and we define
\begin{equation}
\label{deftheta} \theta \equiv \exp \left( \frac{2\pi i}{M} \right)
\end{equation}

If we place $N$ coincident D3-branes at the orbifold singularity of
the $\C^2 / \Z_M \times \C$ orbifold we get a correspondence between
a four-dimensional $\CN = 2 $ quiver gauge theory, which we here dub
the $\CN=2$ {\sl orbifolded quiver gauge theory}, and type IIB
string theory on $\ads_5 \times S^5/ \Z_M$ \cite{Douglas:1996sw}.

$\CN=2$ orbifolded quiver gauge theory consists of $M$ vector
multiplets and $M$ hypermultiplets. Thus, the gauge group of the
$\CN=2$ orbifolded quiver gauge theory is $M$ products of $U(N)$,
one of each node in the quiver. The easiest way to realize $\CN=2$
orbifolded quiver gauge theory is to start with $\CN=4$ SYM theory
with gauge group $U(NM)$, and then do a $\Z_M$ projection. The
surviving components of the scalar fields  are $N\times N$ matrices
which are embedded in $MN\times MN$ $\mathcal{N}=4$ variables as
follows

\begin{eqnarray}\label{ZXW}
&&Z = \pmatrix{0 & Z_1 &  & & & \cr
                       &  0& Z_2 & &  & \cr
                        & & & & \ddots&\cr
                        &  &   &  & 0&Z_{M-1}\cr
                       Z_M & & & & &0\cr}
                        ~,~~
X = \pmatrix{ 0 &  &  & & & X_M\cr
                        X_1& 0 & &  & &\cr
                         &X_2&0& & &\cr
                        &  &   &  & \ddots&\cr
                        & & & &X_{M-1} &0\cr}~,\cr
&&W =\pmatrix{  W_1 & &   &  &  \cr
                        & W_2 & &  &  \cr
                        & &   & \ddots & \cr
                         & & & &W_M \cr}
                                              \end{eqnarray}

In particular, we can consider this projection in the
$\mathfrak{su}(2)$ sector of $U(NM)$ $\CN=4$ SYM theory. In $U(NM)$
$\CN=4$ SYM theory any single-trace operator in the
$\mathfrak{su}(2)$ sector (containing the scalars $Z$ and $X$) can
be gotten from
\begin{equation}
\tr ( Z^{J_1} X^{J_2} )
\end{equation}
by permuting the $X$'s and $Z$'s (and making linear combination).
After the $\Z_M$ projection a single-trace operator should instead
be gotten from
\begin{equation}
\tr ( S^m Z^{J_1} X^{J_2} )
\end{equation}
with the quantization condition
\begin{equation}
\frac{J_1-J_2}{M} \in \Z
\end{equation}
and where $k=0,1,...,M-1$ and $S$ is the twist matrix defined as
\begin{equation}\label{matrixS}
S = \mbox{diag} ( \theta, \theta^2 ,..., \theta^M )
\end{equation}
with $\theta$ given by (\ref{deftheta}). The scalars $Z$ and $X$ obey the following relations
\begin{equation}
Z S = \theta S Z \spa X S = \theta^{-1} S X
\end{equation}

\subsubsection*{Bethe equations}

Define the operator $S_-$ by $S_- Z = X$ and $S_- X = 0$. Then we
can write a general operator with $K$ impurities as
\begin{equation}
\label{genstate} \CO \equiv \sum_{l_1<l_2< \cdots l_K}
\Psi_{l_1,l_2,...,l_K} S_-^{l_1} S_-^{l_2} \cdots S_-^{l_K} \tr (
S^m Z^J )
\end{equation}
where $S_-^l$ acts on the $l$'th $Z$ in $Z^J$, $i.e.$ the sites in $Z^J$ run from site number $1$ to
$J$.

Just as in $\CN = 4$ supersymmetric Yang-Mills theory~\cite{Beisert:2002ff,Beisert:2003tq}, the
computation of dimensions of the operators of interest to us can be elegantly summarized by the
action of an effective Hamiltonian. The $\CN = 4$ dilatation operator is known explicitly in terms of
its action on fields up to two loop order, and implicitly to three loop
order~\cite{Beisert:2003tq,Beisert:2003jb,Eden:2004ua}. That part which is known explicitly can be
projected, using the orbifold projection, to obtain a dilatation operator for the $\CN = 2$ theory.
Here, we shall be interested in computing dimensions of operators in the scalar $\mathfrak{su}(2)$
sector, so we only retain the parts of the operator which will contribute there. They can be obtained
by simply substituting the matrices (\ref{ZXW})  into the $F$-terms of the $\CN = 4$ operator, namely
the relevant dilatation operator on states of the form (\ref{genstate}) is
\begin{equation}\label{dila} D_{1-{\rm{loop}}}=-\frac{g^2_{\rm Y.M.}}{8
\pi^2}{\rm{Tr}}\left(\left|[Z,X]\right|^2\right)
\end{equation}
where $g_{\rm Y.M.}$ is the Yang-Mills coupling constant. This operator acts on nearest neighbors
either as a permutation or as the identity operator and, due to the twist matrix $S$ (\ref{matrixS})
contained in (\ref{genstate}), the one-loop Hamiltonian (\ref{dila}) can be regarded as that of a
Heisenberg $XXX_{1/2}$ spin-chain with twisted boundary
conditions~\cite{Ideguchi:2004wm,DeRisi:2004bc,Beisert:2005he,Astolfi:2006is}. The Bethe equations
are then
\begin{equation}
\label{mbeteqs} \theta^{-2m} \left(
\frac{u_k+\frac{i}{2}}{u_k-\frac{i}{2}} \right)^J = \prod_{j\neq k}
\frac{u_k - u_j + i}{u_k - u_j - i}
\end{equation}
with the momentum constraint
\begin{equation}
\label{mmomcon} \exp \left\{ 2 i \sum_{k=1}^K \arctan \left(
\frac{1}{2u_k} \right) \right\} = \theta^m
\end{equation}
The dispersion relation is still given by (\ref{oneloopen}).

We now introduce the {\sl orbifolded circular string state}. Taking
the logarithm of the Bethe equations (\ref{mbeteqs}) we get%
\footnote{Here and in the following we are considering the branch of
$\arctan$ with $\arctan(0)=0$.}
\begin{equation}
\label{logmbet} - \frac{2 \pi m}{M} + J \arctan \left(
\frac{1}{2u_k} \right) = \sum_{j\neq k} \arctan \left(
\frac{1}{u_k-u_j} \right)
\end{equation}
In general one can freely add a term $\pi n_k$ to this, $n_k \in \Z$ for each impurity, corresponding
to different branches of the logarithm in (\ref{mbeteqs}). However, for the orbifolded circular
string state we consider the specific case in which all $n_k=0$. Due to the momentum constraint
(\ref{mmomcon}) the logarithm of the constraint is taken to be
\begin{equation}
\label{logmmom} 2 \sum_{k=1}^K \arctan \left( \frac{1}{2u_k} \right)
= \frac{2\pi m}{M}
\end{equation}
Summing over all $k$ in (\ref{logmbet}) we get zero on the right-hand side. Combining this with
(\ref{logmmom}) we obtain the constraint
\begin{equation}
\label{Kcon} K = \frac{J}{2}
\end{equation}
Thus, we are considering a state with an equal number of $X$'s and
$Z$'s, $i.e.$ with $J_1=J_2$.

Setting $M=1$ in the above equations we should get a state in the un-orbifolded theory, $i.e.$ in the
$\mathfrak{su}(2)$ sector of $\CN=4$ SYM. Indeed, it is easily seen that
(\ref{logmbet})-(\ref{logmmom}) becomes equal to (\ref{gbeteq})-(\ref{defmm}) in the case
$\alpha=1/2$ and $r=1$. In particular, we see from (\ref{ncon}) that $n= - 2m$. Thus, for $M=1$ the
orbifolded circular string state reduces to the circular string state with $\alpha=1/2$.

Regarding now the Bethe equations (\ref{gbeteq})-(\ref{defmm}) for the circular string we observe
that setting $\alpha=1/2$ and $r=1$ and making the formal replacement $m \rightarrow m/M$ we get the
equations for the orbifolded circular string (\ref{logmbet})-(\ref{Kcon}). Since in solving equations
(\ref{gbeteq})-(\ref{defmm}) one does not use that $m$ is an integer we can thus get the solution for
the orbifolded circular string state merely by making the replacement $m \rightarrow m/M$ in the
solution (\ref{finE}). This reveals that the one-loop energy for the orbifolded circular string state
is
\begin{equation}
\label{Eorbstr} \CE = \frac{\lambda' J^2 m^2}{2 M^2 (J-1)} -
\frac{\pi^2}{6} \frac{\lambda' J^2 m^4}{M^4(J-1)^2} + \CO ( (m/M)^6
)
\end{equation}
for $J \gg 1$. We can furthermore infer from the above considerations that the result
(\ref{prevresult}) for the circular string one-loop energy to order $\lambda'/J$ can be extended to
give the full one-loop energy for the orbifolded circular string state to order $\lambda'/J$
\begin{equation}
\label{Eorbstr2} \CE = \frac{\lambda' J m^2}{2 M^2} + \frac{\lambda'
m^2}{2 M^2} + \frac{\lambda'}{2} \sum_{n=1}^\infty \left[ n^2
\sqrt{1-\frac{4m^2}{n^2 M^2} } -n^2 + \frac{2m^2}{M^2} \right] +
\CO( \lambda' J^{-1} )
\end{equation}

We can now introduce the {\sl winding state} as the orbifolded circular string state in the limit of
large $J$ and large $M$ with $J/M^2$ fixed. To obtain the energy from that of the orbifolded circular
string state (\ref{Eorbstr}) we should consider how this limit affects the derivation of the energy
in Section \ref{sec:bethesolution}. Due to the replacement $m \rightarrow m/M$ we see that the
somewhat formal expansion (\ref{mexpansion}) in $m$ that we introduced in Section
\ref{sec:bethesolution} is replaced by the expansion
\begin{equation}
\label{mMexpansion} f(J,\alpha,m) = \frac{m^2}{M^2} g_1(J,\alpha) +
\frac{m^4}{M^4} g_2(J,\alpha) + \frac{m^6}{M^6} g_3(J,\alpha) +
\cdots
\end{equation}
which, for $M \gg m$, is a perturbative expansion in $m/M$. Therefore, the steps in the derivation of
the energy in Section \ref{sec:bethesolution} remain valid also in the limit $M \rightarrow \infty$.
In fact, the regime in which the derivation in Section \ref{sec:bethesolution} is valid is larger for
the winding state since in the large $M$ limit the non-perturbative effects are negligible. This is
most easily seen by considering (\ref{Eorbstr2}) from which one can see that for $M \rightarrow
\infty$ we can ignore completely the non-perturbative contributions to the energy. In detail, from
(\ref{Eorbstr}) we see that the one-loop energy of the winding state is
\begin{equation}
\label{Ewin1} \CE = \lambda' \frac{\tilde{m}^2}{2 (1-1/J)} -
\lambda' \frac{\pi^2 \tilde{m}^4}{6(J-1)^2} + \CO ( (m/M)^6 )
\end{equation}
for $J \gg 1$ and where we defined
\begin{equation}
\tilde{m} \equiv \frac{\sqrt{J}}{M} m \label{mtilde}
\end{equation}
which is fixed in the limit $M,J \rightarrow \infty$. Expanding in $1/J$, we get from (\ref{Ewin1})
\begin{equation}
\label{Ewin2} \CE = \frac{\lambda' \tilde{m}^2}{2} \left[ 1 +
\frac{1}{J} + \frac{1}{J^2} + \frac{1}{J^3} - \frac{\pi^2
\tilde{m}^2}{3} \left( \frac{1}{J^2} + \frac{2}{J^3} \right) +
\frac{C \tilde{m}^4}{J^3} + \CO( J^{-4} ) \right]
\end{equation}
where $C$ is a constant which is not determined from (\ref{Ewin1}). From this expression it is clear
that for each order in $1/J$ we only have a finite number of terms in powers of $m$. Thus, as
advertised in Section \ref{sec:bethesolution}, we see that for the winding state we can find the
expression for the energy for $m=1$ by making an expansion in $m/M$, since for each power of $1/J$ we
only have a finite number of terms in the $m/M$ expansion.

Finally, we can combine the result (\ref{Ewin2}) with the result for the energy that one gets from
(\ref{Eorbstr2}) in the large $M$ expansion. In this way we can get the leading contribution to the
energy for each power of $m$. In particular, for $m^6$ this is $-2\zeta(4) \lambda' \tilde{m}^6 /J^3$
while for $m^8$ it is $-5 \zeta(6) \lambda' \tilde{m}^8 /J^4$. Using the $m^6$ term we determine $C =
-4 \zeta(4)$ in (\ref{Ewin2}), revealing the following one-loop energy for the winding state to order
$\lambda'/J^3$
\begin{equation}
\label{Ewin3} \CE = \frac{\lambda' \tilde{m}^2}{2} \left[ 1 +
\frac{1}{J} + \frac{1}{J^2} + \frac{1}{J^3} - \frac{\pi^2
\tilde{m}^2}{3} \left( \frac{1}{J^2} + \frac{2}{J^3} \right)
-\frac{2 \pi^4 \tilde{m}^4}{45 J^3} + \CO( J^{-4} ) \right]
\end{equation}
for $J \gg 1$.

%%%%%%%%%%%%%%%%%%%%%%%%%%%%%%%%%%%%%%%%%%%%%%%%%%%%%%%%%%%%%
\subsection{Winding state from Landau-Lifshitz sigma-model}
\label{sec:winLL}

We now consider the $\CN=2$ orbifolded supersymmetric quiver gauge theory version of the LL model. To
obtain the sigma-model that describes it one should use the one-loop Hamiltonian for the $\CN=2$
orbifolded supersymmetric quiver gauge theory acting on operators of the form (\ref{genstate}).
However, since this Hamiltonian corresponds to the one of $XXX_{1/2}$ Heisenberg spin-chain with
twisted boundary conditions, the only difference with respect to the $\CN=4$ SYM is that $\varphi$ is
not periodic in $2\pi$ but instead it has the periodicity
\begin{equation}
\label{s2id}
\varphi \equiv \varphi + \frac{4\pi}{M}
\end{equation}
Thus, the sigma model action is still given by eq.~(\ref{II}), but with the $S^2$ target space
replaced by an $S^2$ with the identifications (\ref{s2id}).

Before studying fluctuations around the winding mode in the orbifolded LL model we first briefly
consider the case without winding mode $m=0$ and without the orbifolding $M=1$. In this case we
quantize the LL action by zooming in on the point $(\theta,\varphi)=(0,0)$ (corresponding to the
point $\vec{n} = (1,0,0)$). Thus, we have that the $S^2$ target-space becomes $\R^2$ in this limit.
As found in \cite{Minahan:2005mx,Minahan:2005qj} a convenient parametrization of $S^2$ for studying
these fluctuations is in the coordinates $z_1,z_2$ defined in eq.~(\ref{minapara}) in
Appendix~\ref{app:twosphere}. This parametrization is particularly useful in that the Hamiltonian
only consists of terms with an equal number of annihilation and creation operators. This made the
computations at order $1/J^2$ in \cite{Minahan:2005mx,Minahan:2005qj} possible. We therefore use this
parametrization of $S^2$ below to study the fluctuations around the winding mode.

We want to study quantum corrections to the classical energy of the
winding state which we denote $|m\rangle$. Classically, a winding
mode corresponds to
\begin{equation}
\label{claswin} \theta = 0 \spa \varphi = \frac{2m}{M} \sigma
\end{equation}
We want to consider quantum fluctuations around this state using the parametrization (\ref{minapara})
of the two-sphere. This can be done using the relation eq.~(\ref{transfo}) between the variables
$(\theta,\varphi)$ and the $(z_1,z_2)$ coordinates for the two-sphere. We therefore write
\begin{equation}\label{varphi}
\varphi = \frac{2m}{M} \sigma + \arcsin \left( \frac{2 f}{\sqrt{J}}
\sqrt{ \frac{ 1-\frac{f^2+g^2}{J}}{1-\frac{4g^2}{J} \left( 1 -
\frac{f^2+g^2}{J} \right) }} \right)
\end{equation}
\begin{equation}\label{theta}
\sin \theta = \frac{2g}{\sqrt{J}} \sqrt{1-\frac{f^2+g^2}{J}}
\end{equation}
This is found from (\ref{transfo}) by substituting $z_1=f/\sqrt{J}$ and $z_2=g/\sqrt{J}$ and by
adding the winding to the $\varphi$ coordinate.

In the following we consider the limit $J, M \rightarrow \infty$ with $J/M^2$ fixed. In this limit we
get that the target space, which is a two-sphere with identifications (\ref{s2id}), becomes a
cylinder $S^1 \times \R$ with $f$ parameterizing the circle of radius $\sqrt{J}/M$ and $g$
parameterizing $\R$.

The Lagrangian can be written in a $1/J$  expansion as
\begin{equation}\label{fglag}
J \CL = 4 g \dot{f} - \CH_2 - \frac{1}{J} \CH_4 -
\frac{1}{J^2} \CH_6 + \cdots
\end{equation}
with
\begin{equation}\label{freeH}
\CH_2 = \left( f' + \tilde m \right)^2 + (g')^2
\end{equation}
where $\tilde m$ is defined in (\ref{mtilde}). We see that the leading order Hamiltonian
(\ref{freeH}) correctly corresponds to having a winding mode with winding number $m$ around the
circle direction of the cylinder $S^1\times \R$ with radius $\sqrt{J}/M$. The linearized EOMs are
\begin{equation}
\dot f=-\frac{1}{2}g'', ~~~~~~\dot g=\frac{1}{2}f''
\label{circeom2}
\end{equation}
Their solution can be written in the following way
\begin{equation}
f = \frac{1}{2} ( A + A^\dagger ) \spa g = \frac{i}{2} ( A^\dagger -
A ) \label{fgwind}
\end{equation}
where we introduced
\begin{equation}
A=\sum_{n\neq 0} a_n e^{-i\frac{n^2}{2} t+in\sigma} \label{bigA}
\end{equation}

Turning to the $1/J$ and $1/J^2$ corrections in the Lagrangian (\ref{fglag}), we find from a
classical computation the following terms
\begin{equation} \label{ch4}
\CH_4 = 2 (f f' + g g')^2 - (f^2+g^2)((f')^2+(g')^2) +
\left(f'f^2-5f'g^2+6fgg' \right)\tilde m - 4\tilde m^2 g^2
\end{equation}
\begin{eqnarray} \label{ch6}
\CH_6 =&& (f^2+g^2) (f f' + g g')^2 +
\left(g(f^2+g^2)(gf'-fg')+\frac{3}{4}(f^2+g^2)^2f'\right) \tilde m
\cr &&+ 2\tilde m^2(2g^4 + (fg)^2+ (gf)^2)
\end{eqnarray}
up to total derivatives with respect to the time $t$. Note that the
$m$-independent terms match those found in
\cite{Minahan:2005mx,Minahan:2005qj}. As usual, the Hamiltonian is
\begin{equation}
H = \frac{\lambda'}{4\pi } \int_0^{2\pi} d\sigma \left( \CH_2 + \frac{1}{J} \CH_4 + \frac{1}{J^2}
\CH_6 + \cdots  \right)
\end{equation}
Now we can treat these Hamiltonians as perturbations of the free Hamiltonian (\ref{freeH}) and
compute the energy of the winding state perturbatively.

In the following we want to use the parts of (\ref{ch4})-(\ref{ch6}) proportional to $m^2$ to compute
the corrections to the energy of the winding state. As part of this, we need to consider the vacuum
expectation value of $\CH_4$ and $\CH_6$. As we discuss below, this is trivially zero for the
$m$-independent terms and the terms proportional to $m$ since they contain derivatives with respect
to $\sigma$. However, for the terms proportional to $m^2$ the expectation value can depend on the
ordering of the $f$ and $g$ variables. Their commutation relations are given in eq.~(\ref{comm}). In
particular, the absence of the zero mode in~(\ref{fgwind}), which gives rise to the $-i/2$ term in
eq.~(\ref{comm}),
plays a crucial role in our calculation.%
\footnote{There is no zero-mode since it cannot correspond to a
physical excitation. The absence of the zero-mode means that we are
instead in a sector of fixed total spin $S_z = - i
\partial_\varphi = (J_1-J_2)/2$. See also \cite{Bertolini:2002nr,windingstring}.} The
quantum operators corresponding to (\ref{varphi}) and (\ref{theta}) can be consistently defined in an
expansion for large $J$.  Using $\zeta$-function regularization one can show that, even at the same
point $\sigma$, $f$ and $g$ do not commute and they have the following commutation relation
\begin{equation}\label{commsp}
[f(t,\sigma),g(t,\sigma)]=-\frac{i}{2}
\end{equation}
Consequently $f$ and $g$ can be treated as non-commuting coordinates of a 2 dimensional space and one
can use the general procedure of non-commutative geometry, $i.e.$ the Weyl prescription, for
associating a quantum operator to a classical function~\cite{Szabo:2001kg}. This technique provides a
systematic way to describe noncommutative spaces in general. Although we will focus solely on the
commutator (\ref{commsp}), Weyl quantization also works for more general commutation relations.

Let us consider the commutative algebra of functions on a 2 dimensional Euclidean space $\R^2$, with product defined by the
usual multiplication of functions. In general a function $F(f,g)$ may be described by its Fourier transform
\begin{equation}\label{fourier}
\tilde F(k_f,k_g) = \int df dg e^{-i k_f f-ik_g g} F(f,g)
\end{equation}
with  $\tilde F(-k) = \tilde F(k)$ whenever $F(x)$ is real-valued. We define a noncommutative space
by replacing the local coordinates $f$ and $g$ of $\R^2$ by Hermitian operators $\hat f$ and $\hat g$
obeying the commutation relations~(\ref{commsp}). The $\hat f$ and $\hat g$ then generate a
noncommutative algebra of operators.

Given the function $F(f,g)$ and its corresponding Fourier coefficients (\ref{fourier}), we introduce
its Weyl symbol by
\begin{equation}\label{weylsymb}
W [F] = \int \frac{dk_f dk_g}{(2\pi)^2} \tilde F(k) e^{ik_f\hat f+i k_g \hat g} ,
\end{equation}
where we have chosen the symmetric Weyl operator ordering prescription. The Weyl operator $\hat W
[F]$ is Hermitian if $F(f,g)$ is real-valued. We can write (\ref{weylsymb})  in terms of an explicit
map $\hat M (\hat f,\hat g)$ between operators and fields by using (\ref{fourier}) to get
\begin{equation}\label{weylop}
\hat W [F] = \int df dg F(f,g) \hat M (\hat f,\hat g)
\end{equation}
where
\begin{equation}\label{qoperator}
   \hat M(\hat f,\hat g)=\int \frac{dk_f dk_g}{(2\pi)^2}e^{i k_f\hat f+i k_g \hat g} e^{-i k_f f-i k_g  g}
\end{equation}
The operator (\ref{qoperator}) is Hermitian, $\hat M^\dagger=\hat M$, and it describes a mixed basis
for operators and fields on spacetime. In this way we may interpret the field $F(f,g)$ as the
coordinate space representation of the Weyl operator $\hat W [F]$. Note that in the commutative case,
the map (\ref{qoperator}) reduces trivially to a delta-function. But generally, by the
Baker-Campbell-Hausdorff (BCH) formula, it is a highly non-trivial field operator. Using
(\ref{commsp}) the BCH formula gives
\begin{equation}\label{BCH}
e^{i k_f\hat f+i k_g \hat g}=e^{i k_f\hat f}e^{i k_g \hat g}e^{-\frac{i}{4} k_f k_g}
\end{equation}
When this is inserted in (\ref{weylsymb}) one can construct explicitly the quantum operator
corresponding to any classical polynomial in $g$ and $f$. Consider for example eq.(\ref{theta}),
expanding for large $J$ we get
\begin{equation}\label{thetaexp}
\sin \theta = \frac{2g}{\sqrt{J}}\left[1-\frac{f^2+g^2}{J}-\frac{(f^2+g^2)^2}{4J^2}+
\CO\left( J^{-3} \right)\right]
\end{equation}
Using the definitions above the corresponding Weyl ordered operator (\ref{weylop}) reads
\begin{eqnarray}
\label{thetaexpop} W[ \sin \theta ] &=& \frac{2\hat g}{\sqrt{J}}-\frac{1}{J^{3/2}}\left[(\hat
f^2+\hat g^2)\hat g+\frac{i}{2}\hat f\right]\cr&& +\frac{i}{4 J^{5/2}}\left[i(\hat f^2+\hat
g^2)^2\hat g-\hat f(\hat f^2+\hat g^2)- \frac{i}{4}\hat g\right]+\CO\left( J^{-7/2}\right)
\end{eqnarray}
We can therefore construct explicitly the Hamiltonian densities appearing in (\ref{fglag}) which are
relevant at each order in the $1/J$ expansion. The terms in the Hamiltonian proportional to $m^2$ are
$\tilde{m}^2 \cos^2 \theta $ which thus become
\begin{equation}
\tilde{m}^2 (1 - W[ \sin \theta ]^2) = \tilde{m}^2 \left[ 1 -
\frac{4}{J} \hat{g}^2 + \frac{2}{J^2} ( 2\hat{g}^4 +
(\hat{f}\hat{g})^2+(\hat{g}\hat{f})^2 ) + \CO( J^{-3} ) \right]
\end{equation}
Upon removing the hats we get precisely the $m^2$ parts in (\ref{ch4}) and (\ref{ch6}).

Since the winding state $ | m \rangle$ is our vacuum state we require it to satisfy
\begin{equation}
A | m \rangle = 0
\end{equation}
From the commutator (\ref{commsp}) we see that $A$ and $A^\dagger$ have the commutator (at the same
point)
\begin{equation}
[A , A^\dagger] = -1
\end{equation}
Using this, we obtain that the vacuum state $ | m \rangle$ satisfies
the following properties
\begin{equation}
\langle m | A A^\dagger | m \rangle = -1 \spa
\spa \langle m | (A A^\dagger)^2 | m \rangle = 1  \spa \langle m | A^2
(A^\dagger)^2 | m \rangle = 2
\end{equation}
In addition, we have that any vacuum expectation value involving $A'$ and $A'^\dagger$ is zero.

The Hamiltonian densities can then be written as
\begin{equation}
\CH_2 = \frac{1}{2} ( A' A'^\dagger + A'^\dagger A' ) + (A' +
A'^\dagger) \tilde m + \tilde m^2
\end{equation}
\begin{equation}
\CH_4 = (\cdots) + (\cdots) \tilde m+ ( A^2 + (A^\dagger)^2 - A
A^\dagger - A^\dagger A   ) \tilde m^2
\end{equation}
\begin{equation}
\CH_6 = (\cdots) + (\cdots) \tilde m + ( \cdots + \frac{1}{2} A^2
(A^\dagger)^2 ) \tilde m^2
\end{equation}
where in $\CH_4$ and in $\CH_6$ we only wrote the relevant terms that can contribute to the
computation of the energy corrections of the winding state.

The classical value of the energy of the winding state is given by
\begin{equation}
\CE_0=\langle m | H_2 | m \rangle =\frac{ \lambda'\tilde m^2 }{2}
\end{equation}
The $1/J$ correction to the classical energy is
\begin{equation}
\label{EE1} \CE_1=\langle m |H_4 | m \rangle = \frac{\lambda'\tilde
m^2 }{2J}
\end{equation}
We then go on to compute the $1/J^2$ correction. This is given by
the second order perturbation theory formula
\begin{equation}
\CE_2=\sum_{|i\rangle \neq |m\rangle}\frac{\big|\langle
m|H_4|i\rangle \big|^2}{\CE_{m}-\CE_i}+\langle m|H_6|m\rangle
=\CE_A+\CE_B \label{secwin}
\end{equation}
where $|i\rangle$ is an intermediate state. The computation of
$\CE_B$ is immediately done and we get
\begin{equation}
\CE_B=\langle m | H_6 | m\rangle = \frac{\lambda'\tilde m^2
}{2J^2}\label{EB}
\end{equation}
We now compute $\CE_A$. It is not difficult to see that the only
possibility for obtaining a non-zero result is that the intermediate
state $|i\rangle$ is of the form
\begin{equation}
| i\rangle \equiv a_n^{\dagger} a_{-n}^\dagger | m \rangle
\end{equation}

We compute $[[ A^2 |_{t=0} , a_n^\dagger ], a_{-n}^\dagger ] = 2$ and from this we get $ [[ \CH_4 ,
a_n^\dagger ], a_{-n}^\dagger ] =  2 \tilde m^2 $. Thus we have
\begin{equation}
\langle m | \CH_4 | i \rangle = \langle m | \CH_4 a_n^{\dagger}
a_{-n}^\dagger | m \rangle = \langle m | [[ \CH_4 , a_n^\dagger ],
a_{-n}^\dagger ] | m \rangle = 2\tilde m^2
\end{equation}
where $n \neq 0$. Moreover we have that $\CE_m = \tilde m^2 $ and
that the intermediate state $|i\rangle$ has energy $\CE_i = \tilde
m^2 + 2 n^2$. Therefore we have
\begin{equation}
\CE_A = \sum_{n =1}^\infty \frac{|\langle m | H_4 | i
\rangle|^2}{\CE_m - \CE_i} = - \frac{4\lambda'\tilde m^4}{J^2}
\sum_{n=1}^\infty \frac{1}{n^2} = - \frac{2\lambda'\tilde
m^4\pi^2}{3 J^2}
\end{equation}

Up to this order we can write the corrected energy of the winding
state as
\begin{equation}
\label{Estring} \CE = \frac{\lambda'}{2} \tilde m^2\left[ 1 +
\frac{1}{J} + \frac{1}{J^2} \left( 1 - \frac{\pi^2 }{3}\tilde m^2
\right) + \CO( J^{-3} ) \right]
\end{equation}
We note that in the orbifold case, contrary to the usual case, there is no higher derivative term
that has to be included up to this order. We see that eq.~(\ref{Estring}) precisely matches the
result obtained from the Bethe ansatz (\ref{Ewin3}). If we set $M=1$, $r=1$ and $\alpha=1/2$, it also
matches the Bethe ansatz result (\ref{finEexp}).

It is interesting to test whether our results above relies on choosing the Weyl ordering for $f$ and
$g$ in (\ref{thetaexpop}). More generally, we can take the classical expression (\ref{thetaexp}) and
try the most general way to multiply $g$ with $f^2+g^2$ while retaining the same classical limit.
This is parameterized as the following more general ordering prescription for (\ref{thetaexp})
\begin{eqnarray}
\label{altord} && W_{\rm gen} [\sin \theta ] = \frac{2\hat{g}}{\sqrt{J}} -  \frac{(1-a_1)
\hat{g}(\hat{f}^2+ \hat{g}^2) + a_1 (\hat{f}^2+\hat{g}^2)\hat{g}}{J^{3/2}} \cr&&
 - \frac{(1-a_2-a_3) \hat{g}(\hat{f}^2+\hat{g}^2)^2 + a_2 (\hat{f}^2+\hat{g}^2)\hat{g}(\hat{f}^2+\hat{g}^2)
  + a_3(\hat{f}^2+\hat{g}^2)^2 \hat{g}}{4 J^{5/2}} + \CO( J^{-7/2} )\cr&&
\end{eqnarray}
The Weyl ordering corresponds to the particular choice $a_1=1/2$, $a_2=-1/2$ and $a_3=3/4$. Clearly,
using this ordering cannot affect the $-4\tilde{m}^2\hat{g}^2/J$ term at order $1/J$ which gives rise
to the $m^4/J^2$ correction to the energy of the winding state. However, for the $m^2/J^2$
correction, which comes from the expectation value $\langle m | H_6 | m\rangle$, it could have an
effect. Using (\ref{altord}) we compute
\begin{equation}
\tilde{m}^2 (1 - W_{\rm gen}[ \sin \theta ]^2) = \tilde{m}^2 \left[
1 - \frac{4}{J} \hat{g}^2 + \frac{2}{J^2} \left[ 2\hat{g}^4 +
2(\hat{f}\hat{g})^2+ i a_1 (\hat{f}\hat{g}+\hat{g}\hat{f}) \right] +
\CO( J^{-3} ) \right]
\end{equation}
Computing now the expectation value $\langle m | H_6 | m\rangle$ we find that it is equal to
(\ref{EB}) for any value of $a_1$, $a_2$ and $a_3$. Thus, amazingly, the expectation value $\langle m
| H_6 | m\rangle$ is independent of the choice of ordering prescription. This makes it a rather solid
prediction.

\subsubsection*{Higher orders}

Going to the next order, namely $\lambda'/J^3$, is a non-trivial
task and one has first to construct the quantum Hamiltonian $\CH_8$
using the Weyl prescription and then use perturbation theory to the
next order. The quantum expression for $\CH_8$ and the perturbative
calculations are too complicated to display here, but for the term
proportional to $\tilde m^2$ they give%
\footnote{Note here that one can find values of $a_1$, $a_2$ and $a_3$ such that the expectation
value of the $1/J^3$ correction to the Hamiltonian gives $\frac{\lambda'}{2J^3} \tilde m^2$ which is
the result predicted by the gauge theory (\ref{Ewin3}). However this does not seem the correct way to
resolve the discrepancy.}
\begin{equation}\label{m2j3}
\frac{\lambda'}{8J^3} \tilde m^2
\end{equation}
This, as expected, is not in agreement with the Bethe ansatz result (\ref{Ewin3}). To get to
(\ref{m2j3}) one has to use $\zeta$-function regularization to define the sums over intermediate
states in perturbation theory and this regularization, as was argued in Ref.~\cite{Minahan:2005mx},
is not expected to give the correct results at this order. The same arguments apply for the term
proportional to $\tilde m^4$ where, moreover, in the sigma model action (\ref{II}) one should include
also the higher derivative term in (\ref{ham2}). More precisely, we should add the following term to
the Hamiltonian
\begin{equation}
H_{h.d.} = \frac{J\lambda'}{4\pi} \int_0^{2\pi} d\sigma \left[ - \frac{\pi^2}{12 J^2} (\vec{n}'')^2 +
\CO (J^{-4}) \right]
\end{equation}
Evaluated on the winding state this Hamiltonian gives
\begin{equation}
\langle m|H_{h.d.}|m \rangle = - \frac{2\lambda'\pi^2 }{3 J^3}\tilde
m^4
\end{equation}
Putting all together we do not get a result which is consistent with the solution
of the Bethe equations. It is however possible to provide an ordering prescription,
which differs from the Weyl ordering only starting from the order $\lambda'/J^3$,
and that gives instead results for the $\lambda'/J^3$ terms that match precisely
those of the Bethe equations.

%%%%%%%%%%%%%%%%%%%%%%%%%%%%%%%%%%%%%%%%%%%%%%%%%%%%%%%%%%%%%
\sectiono{Conclusions}
\label{sec:concl}

The main goal of this paper has been to investigate how well the
continuous Landau-Lifshitz sigma model reproduces the results
obtained from the Bethe equations that provide the one-loop scaling
dimensions of operators in the $\mathfrak{su}(2)$ sector of $\CN=4$
SYM. This has been examined by computing the finite size corrections
to the energy of the circular rotating string and to its orbifolded
generalization. In particular we also considered the winding state
which is obtained as the limit of the orbifolded circular string
solution when $J\to\infty$ and $J/M^2$ is kept fixed.

Our conclusion is that, in the case of the circular rotating string, for the leading $1/J$
corrections in $m^2$ and $m^4$ and for the subleading $1/J^2$ corrections to the $m^2$ term, we
found, in fact, complete agreement between the perturbative LL energies and the solutions of the
Bethe equations. In the case of the winding state we obtain full agreement of the finite-size
corrections up to, and including, the $1/J^2$ order. In particular, the matching of the $m^4/J^2$
term represents an important example of a successful match of gauge theory and a continuous
sigma-model which relies on summing over the so called ``impurity non preserving channels'', namely
intermediate states excited by a number of oscillators different from the external state.

The Bethe equations have been solved by using a novel procedure
which allows to determine the finite size corrections to the
dimensions of these operators and the energies of the corresponding
string states in an expansion in the winding number $m$ but to all
orders in the angular momentum $J$. We also applied the same
procedure to solve the Bethe equations for the $XXX_{-1/2}$
Heisenberg spin chain, which instead describe the one-loop
contribution to the $\mathfrak{sl}(2)$ sector. According to the AFS
ansatz~\cite{Arutyunov:2004vx} the solution in this case should
describe, on the string side of the AdS/CFT duality, the spinning
string.

We were then able to match the results of the $\mathfrak{su}(2)$
sectors with those derived from an effective quantum LL model, both
in its orbifolded and un-orbifolded versions. The matching is highly
non-trivial since it requires in the LL model second order
perturbation theory and $\zeta$-function regularization.

The calculations of quantum finite-size corrections are much simpler
in the LL model than in the full superstring computation since in
this model one does not include the contributions of the bosonic and
fermionic modes  which do not belong to the $\mathfrak{su}(2)$
sector. Omitting these and other string modes is obviously  not
correct in general, but in some simple cases it may happen that the
role of these extra modes may be just to provide a particular UV
regularization of the quantum LL result. This was first suggested in
\cite{Beisert:2005mq} and it is by now established that the
$\zeta$-function regularization is the correct one for reproducing
the first order finite size correction. Our results for the circular
rotating string and for the winding state confirm this observation.
However the $\zeta$-function regularization might not be the
appropriate one when comparing the results of the next to leading
order finite size corrections, as suggested also
in~\cite{Minahan:2005mx,Minahan:2005qj}. The mismatch that we found
for the $m^4/J^2$ term in the case of the circular rotating string,
and for the $m^2/J^3$ term in the case of the winding state, clearly
point towards the necessity of a full superstring calculation that
would naturally suggest the correct regularization prescription and
hopefully provide complete agreement with the Bethe ansatz results.

In a subsequent paper~\cite{windingstring} we shall show that, for the pp-wave background studied
in~\cite{Bertolini:2002nr}, the superstring degrees of freedom which are relevant at the first order
in $\lambda'$ are precisely those of the $ \mathfrak{su}(2)$ sector and, consequently, the
superstring sigma model reduces to the LL model in the particular set of coordinates used here to
study the winding state. In this context we shall also compute directly from the superstring theory
the $1/J^2$ corrections to the string excitations. For the winding state we shall show that for the
$m^2/J$ term all the divergences in the LL model, which we need to regularize, cancel in the
superstring calculation when including the zero-mode contributions from the full superstring theory,
including fermions and all the transverse directions, and the finite piece that remains is precisely
(\ref{EE1}). This strongly suggests that the full superstring theory picks up a particular
regularization prescription which allows to match the gauge theory side.

There are some interesting extensions of this work. An obvious generalization is the analysis of the
$\mathfrak{sl}(2)$ sector of operators of the type $\tr ( D^{s_1} Z D^{s_2} Z \cdots D^{s_J} Z )$.
The spin chain LL action  also in this case should match the string
action~\cite{Stefanski:2004cw,Bellucci:2004qr} and consequently a matching of the higher order finite
size corrections between the solution of the Bethe equations (\ref{sl2en}) and the LL results should
be easily obtained along the lines of the present paper.

It would be interesting to study also the $\beta$-deformed version of AdS/CFT which relates an
exactly marginal superconformal deformation of SYM theory to string theory in the $AdS_5\times
(S^5)_\beta$ background constructed by applying a combination of $T$-duality, shift of angle and
another $T$-duality to the original $AdS_5\times S^5$ background~\cite{Lunin:2005jy}. The existence
of integrable structures on the two sides of the duality was first discussed in~\cite{Frolov:2005ty}
where it was argued that the integrability of strings in $AdS_5\times S^5$ implies the integrability
of the deformed world sheet theory with real deformation parameter. The Bethe equations are identical
to those solved here for the orbifold case. The $\beta$-deformed Bethe equations in fact contain a
twist as in the orbifold but with $2 m/M\to \beta J$. Also in this case one can focus on the
solutions of the Bethe equations with equal mode numbers for all the roots and compute the higher
order finite size corrections with the procedure developed in this paper. The results can then be
compared with one loop string theory results and with possible LL calculations~\cite{Frolov:2005ty}.
This might lead to a better understanding of the spectrum of strings in less-supersymmetric
backgrounds.

Finally, it would be important to study if the procedure used in
this paper to solve the Bethe equations could be extended also to
higher powers of the winding number of the string states.

\section*{Acknowledgments}

We thank N. Gromov, J. A. Minahan, A. Tirziu and A. A. Tseytlin for useful discussions.

%%%%%%%%%%%%%%%%%%%%%%%%%%%%%%%%%%%%%%%%%%%%%%%%%%%%%%%%%%%

\begin{appendix}
\addcontentsline{toc}{section}{Appendices}
%%%%%%%%%%%%%%%%%%%%%%%%%%%%%%%%%%%%%%%%%%%%%%%%%%%%%%%%%%%%%
\sectiono{Zeroes of the Laguerre polynomial} \label{app:laguerre}

The Laguerre differential equation is
\begin{equation}
\label{lagpol} \left[ y \frac{d^2}{dy^2} + ( \nu + 1 -y)
\frac{d}{dy} + \mu \right] f(y) = 0
\end{equation}
The associated Laguerre polynomial, which is a solution of (\ref{lagpol}), is written
$f=L^\nu_\mu(y)$.

We write the zeroes of the Laguerre polynomial as $y_k$,
$k=1,2,...,\mu$, where we here used the general property of the
Laguerre polynomial that the number of zeroes equals $\mu$. For the
zeroes of the Laguerre polynomial we have the sum rules
\cite{Ahmed:1978uw}
\begin{equation}
\label{zerosum1} y_k \sum_{j=1, j \neq k}^{\mu} (y_k - y_j)^{-1} = -
\frac{1}{2} \big[ 1 + \nu - y_k \big]
\end{equation}
\begin{equation}
\label{zerosum3} y_k^3 \sum_{j=1, j \neq k}^{\mu} (y_k - y_j)^{-3} =
- \frac{1}{8} \big[ (\nu+1)(\nu+3) - (2\mu+\nu+1) y_k \big]
\end{equation}

%%%%%%%%%%%%%%%%%%%%%%%%%%%%%%%%%%%%%%%%%%%%%%%%%%%%%%%%%%%%%%%%%

\sectiono{Circular rotating string} \label{appB}

In this Appendix we repeat the computation of Sec.~\ref{circular}
using a different coordinate system.

We consider again fluctuations around
\begin{equation}
\label{circ} \theta = 0 \spa \varphi = 2m\sigma
\end{equation}
for which the classical energy is
\begin{equation}
E_0=J\left(1+\lambda' \frac{m^2}{2}+\mathcal{O}(\lambda'^2)\right)
\label{clascirc}
\end{equation}
We want to compute quantum corrections to the energy of the ground state $|0_m\rangle$. In Section
\ref{circular} we used the parametrization (\ref{spherical}) for the two-sphere in doing this. Here
we instead choose to consider the Landau-Lifshitz action using the parametrization (\ref{minapara}).
Using (\ref{transfo}) we see that this means we should consider the following form for the
fluctuations around the solution (\ref{circ})
\begin{equation}
\varphi = 2m \sigma + \arcsin \left( \frac{2 f}{\sqrt{J}}
\sqrt{ \frac{ 1-\frac{f^2+g^2}{J}}{1-\frac{4g^2}{J} \left( 1 -
\frac{f^2+g^2}{J} \right) }} \right)
\end{equation}
\begin{equation}
\sin \theta = \frac{2g}{\sqrt{J}} \sqrt{1-\frac{f^2+g^2}{J}}
\end{equation}
where $m$ is the winding number. The Lagrangian is given again by eq. (\ref{fgcirclag}) but now we
have the following identification
\begin{eqnarray}
&&\CH_2 = ( f')^2 + (g')^2 -4m^2 g^2
\\
&&\CH_3 =  \left(f'f^2-5f'g^2+6fgg' \right)m \label{h3circ}
\\
&&\CH_4 = 2 (f f' + g g')^2 - (f^2+g^2)[(f')^2+(g')^2] - 4 m^2[g^4 + (fg)^2] \label{h4circ}
\end{eqnarray}
The solution to the linearized EOMs is still given by eq.~(\ref{fgcirc}) with the same definition
(\ref{freq}) for $\omega_n$ and $w_n$. As we can see, the quadratic Hamiltonian $H_2$ is the same in
the two parametrization, as expected. Therefore, for the leading correction to the energy of the
ground state, we get the same expression (\ref{E1circ}), after regularizing appropriately the
divergences.

We now move to the next subleading correction to the energy which is
given by a second order perturbation theory expression as in
Sec.~\ref{circular}, namely
\begin{equation}
E_2=\sum_{|i\rangle \neq |0_m\rangle}\frac{\big|\langle
0_m|H_3|i\rangle \big|^2}{E_{0_m}-E_i}+\langle 0_m|H_4|0_m\rangle
=E_2^1+E_2^2 \label{secc}
\end{equation}
where $|i\rangle$ is an intermediate state. Even though the expression of $H_3$ is very different in
the two parametrizations, it is not difficult to see that both give the same result for $\langle
0_m|H_3|i\rangle $. This means that $E_2^{(1)}$ is again given by eq.~(\ref{E21circ}).

The evaluation of $E_2^{(2)}$ is instead slightly different in this
case, therefore we present it in detail. We have
\begin{eqnarray}
&&\frac{\lambda'}{4\pi J}\int_0^{2\pi}d\sigma\left[2 (f f' + g g')^2 -
(f^2+g^2)\left((f')^2+(g')^2\right) \right]\cr &=&\frac{\lambda'}{2 J}\sum_{l,p,q,s=-\infty}^{\infty}
\frac{\sqrt{w_lw_pw_qw_s}}{16}\delta_{l+p+q+s}\,(s-2p)q\cr&& \left[\left(1-\frac{1}{w_lw_p}\right)
\left(1-\frac{1}{w_qw_s}\right)(a_la_pa_{-q}^{\dagger}a_{-s}^{\dagger}+
a_{-l}^{\dagger}a_{-p}^{\dagger}a_qa_s)\right.\cr &&\left.+ \left(1+\frac{1}{w_lw_p}\right)
\left(1+\frac{1}{w_qw_s}\right)( a_{-l}^{\dagger}a_p+a_la_{-p}^{\dagger})(
a_{-s}^{\dagger}a_q+a_sa_{-q}^{\dagger})\right]
\end{eqnarray}
and
\begin{eqnarray}
&&\frac{\lambda'}{4\pi J}\int_0^{2\pi}d\sigma 4m^2\left[g^4+(f g)^2\right]= 2\frac{m^2}{J}
\lambda'\sum_{l,p,q,s=-\infty}^{\infty}\frac{\delta_{l+p+q+s}} {16\sqrt{w_lw_pw_qw_s}}
\cr&&\left[(1+w_lw_q)(a_la_pa_{-q}^{\dagger}a_{-s}^{\dagger}+
a_{-l}^{\dagger}a_{-p}^{\dagger}a_qa_s+a_la_{-p}^{\dagger}a_{-q}^{\dagger}a_s\right.\cr &&\left.+
a_{-l}^{\dagger}a_pa_qa_{-s}^{\dagger})+ (1-w_lw_q)(a_la_{-p}^{\dagger}a_qa_{-s}^{\dagger}+
a_{-l}^{\dagger}a_pa_{-q}^{\dagger}a_s)\right]
\end{eqnarray}
where we only kept terms with two $a$'s and two $a^{\dagger}$'s. We
get
\begin{eqnarray}
&&\langle E_2^2=0_m|H_4|0_m\rangle=\frac{\lambda'}{2J}\sum_{l,p=-\infty}^{\infty}\{\frac{1}{32}
\left\{\left[(p^2+6pl)(w_lw_p+\frac{1}{w_lw_p})\right.\right.\cr &&
\left.\left.-(p^2-2pl)(\frac{w_l}{w_p}+\frac{w_p}{w_l}) -4(p^2+2pl)\right]
+\frac{m^2}{4}\frac{(3+w_l^2)}{w_lw_p}\right\}
\end{eqnarray}
In the previous expression there are divergent sums, some of which
can be regularized as before by adding and subtracting the divergent
contribution and then using $\zeta$-function regularization. We get
\begin{eqnarray}
E_2^2&=&\frac{\lambda'}{2J}\left\{
\frac{1}{16}\sum_{l=-\infty}^{\infty}\sqrt{1-\frac{4m^2}{l^2}}\sum_{p=1}^{\infty}\left[\left(2p\sqrt{p^2-4m^2}
-\frac{p^2}{\sqrt{1-\frac{4m^2}{p^2}}}\right)+4m^2\right]\right.\cr &-&\left.
\frac{1}{16}\left[2\sum_{l=1}^{\infty}\left(\frac{1}{\sqrt{1-\frac{4m^2}{l^2}}}-1\right)-1\right]
\sum_{p=1}^{\infty}\left[\left(2p\sqrt{p^2-4m^2} -\frac{p^2}{\sqrt{1-\frac{4m^2}{p^2}}}\right)
\right.\right.\cr &+&\left.\left.4m^2\right]+
\frac{3m^2}{4}\left[2\sum_{l=1}^{\infty}\left(\frac{1}{\sqrt{1-\frac{4m^2}{l^2}}}-1\right)-1\right]
\left[2\sum_{p=1}^{\infty}\left(\frac{1}{\sqrt{1-\frac{4m^2}{p^2}}}-1\right)-1\right] \right.\cr
&+&\left.\frac{m^2}{4}\sum_{l=-\infty}^{\infty}\sqrt{1-\frac{4m^2}{l^2}}\left[2\sum_{p=1}^{\infty}\left(\frac{1}
{\sqrt{1-\frac{4m^2}{p^2}}}-1\right)-1\right]\right\} \label{e4circ2}
\end{eqnarray}
Here we see that the sum over $l$ in the first and last term on the
r.h.s. of the previous expression is divergent. To obtain a
meaningful result, those divergences need to be removed. We will
show that for the terms proportional to $m^2$ and $m^4$ those
divergences exactly cancel and we conjecture that the same should
happen for all the other terms in the series expansion in powers of
$m^2$. Expanding the result obtained up to this order for the
correction to the energy of the circular string in powers of $m^2$,
up to $m^4$, we obtain
\begin{equation}
E-J=J
\lambda'\left[\frac{m^2}{2}+\frac{1}{J}\left(\frac{m^2}{2}-\frac{\pi^2}{6}m^4+\mathcal{O}(m^6)\right)
+\frac{1}{J^2}\left(\frac{m^2}{2}-\frac{4\pi^2}{3}m^4+\mathcal{O}(m^6)\right)\right]
\label{zeroci}
\end{equation}
where we also included the contribution from the higher derivative term (\ref{classicphd}). We see
that (\ref{zeroci}) coincides with (\ref{zerocirc}) obtained in Sec.~\ref{circular}. It is important
to notice that in (\ref{zeroci}) all the divergent contributions present in (\ref{e4circ2}) which
cannot be regularized, exactly cancel up to the order $m^4$.

%%%%%%%%%%%%%%%%%%%%%%%%%%%%%%%%%%%%%%%%%%%%%%%%%%%%%%%%%%%%%%%%%

\sectiono{Parameterizations of the two-sphere}
\label{app:twosphere}

We use in this paper two different parameterizations of the
two-sphere. The three-dimensional vector $\vec{n} = (n_1,n_2,n_3)$
defines the unit two-sphere as $\vec{n}^2 = 1$. The unit vector
$\vec{n}$ can be parameterized by the spherical coordinates $\theta$
and $\varphi$ as
\begin{equation}
\label{spherical} n_1 = \cos \theta \cos \varphi \spa n_2 = \cos
\theta \sin \varphi \spa n_3 = \sin \theta
\end{equation}
Here $-\pi/2 \leq \theta \leq \pi/2$. Note that in this
parametrization the equator is at $\theta=0$, which corresponds to
$n_3 = 0$. In particular $(\theta,\varphi)=(0,0)$ corresponds to
$\vec{n}=(1,0,0)$. Another parametrization that we use is chosen
such that it is symmetric in exchanging $n_2$ and $n_3$. It was
found in \cite{Minahan:2005mx}. It parameterizes $\vec{n}$ by the
two coordinates $z_1$ and $z_2$ as
\begin{equation}
\label{minapara} n_1 = \sqrt{1-4z^2(1-z^2)} \spa n_2 = 2z_1
\sqrt{1-z^2} \spa n_3 = 2z_2 \sqrt{1-z^2}
\end{equation}
where $z$ is $ z = \sqrt{z_1^2+z_2^2} $.

We see that in this parametrization $(z_1,z_2)=(0,0)$ corresponds to
$\vec{n}=(1,0,0)$. The two above parameterizations
$(\theta,\varphi)$ and $(z_1,z_2)$ are connected through the
following relations
\begin{equation}
\label{transfo} \varphi = \arcsin \left( 2 z_1 \sqrt{
\frac{1-z^2}{1-4z_2^2(1-z^2)} } \right) \spa \sin \theta = 2 z_2
\sqrt{1-z^2}
\end{equation}

%%%%%%%%%%%%%%%%%%%%%%%%%%%%%%%%%%%%%%%%%%%%%%%%%%%%%%%%%%%%%

\end{appendix}

\providecommand{\href}[2]{#2}\begingroup\raggedright\endgroup

%The following two lines is for bibtex only:
%\bibliographystyle{C:/BIB/utphys}
%\bibliography{C:/BIB/mybib,C:/BIB/bibrot}
%\bibliographystyle{../INPUT/utphys}
%\bibliography{../BIB/mybib,../BIB/bibrot}
%\bibliographystyle{BIB/utphys}
%\bibliography{BIB/mybib,BIB/bibrot}

\begin{thebibliography}{10}

\bibitem{Gubser:2002tv}
S.~S. Gubser, I.~R. Klebanov, and A.~M. Polyakov, ``A semi-classical
limit of
  the gauge/string correspondence,'' {\em Nucl. Phys.} {\bf B636} (2002)
  99--114,
\href{http://www.arXiv.org/abs/hep-th/0204051}{{\tt
hep-th/0204051}}.
%%CITATION = HEP-TH 0204051;%%.

\bibitem{Frolov:2002av}
S.~Frolov and A.~A. Tseytlin, ``Semiclassical quantization of
rotating
  superstring in {$\ads_5 \times S^5$},'' {\em JHEP} {\bf 06} (2002) 007,
\href{http://www.arXiv.org/abs/hep-th/0204226}{{\tt
hep-th/0204226}}.
%%CITATION = HEP-TH/0204226;%%.

\bibitem{Frolov:2003qc}
S.~Frolov and A.~A. Tseytlin, ``{Multi-spin string solutions in
{$\ads_5 \times
  S^5$}},'' {\em Nucl. Phys.} {\bf B668} (2003) 77--110,
\href{http://www.arXiv.org/abs/hep-th/0304255}{{\tt
hep-th/0304255}}.
%%CITATION = HEP-TH/0304255;%%.

\bibitem{Minahan:2002ve}
J.~A. Minahan and K.~Zarembo, ``The {Bethe-ansatz} for {$\CN = 4$}
super
  {Yang-Mills},'' {\em JHEP} {\bf 03} (2003) 013,
\href{http://www.arXiv.org/abs/hep-th/0212208}{{\tt
hep-th/0212208}}.
%%CITATION = HEP-TH 0212208;%%.

\bibitem{Beisert:2003tq}
N.~Beisert, C.~Kristjansen, and M.~Staudacher, ``The dilatation
operator of
  {$\CN = 4$} super {Yang-Mills} theory,'' {\em Nucl. Phys.} {\bf B664} (2003)
  131--184,
\href{http://www.arXiv.org/abs/hep-th/0303060}{{\tt
hep-th/0303060}}.
%%CITATION = HEP-TH 0303060;%%.

\bibitem{Beisert:2003yb}
N.~Beisert and M.~Staudacher, ``The {$\CN = 4$} {SYM} integrable
super spin
  chain,'' {\em Nucl. Phys.} {\bf B670} (2003) 439--463,
\href{http://www.arXiv.org/abs/hep-th/0307042}{{\tt
hep-th/0307042}}.
%%CITATION = HEP-TH 0307042;%%.

\bibitem{Beisert:2003ys}
N.~Beisert, ``The {$\mathfrak{su}(2|3)$} dynamic spin chain,'' {\em
Nucl.
  Phys.} {\bf B682} (2004) 487--520,
\href{http://www.arXiv.org/abs/hep-th/0310252}{{\tt
hep-th/0310252}}.
%%CITATION = HEP-TH 0310252;%%.

\bibitem{Mandal:2002fs}
G.~Mandal, N.~V. Suryanarayana, and S.~R. Wadia, ``Aspects of
semiclassical
  strings in ads(5),''
\href{http://arXiv.org/abs/hep-th/0206103}{{\tt hep-th/0206103}}.
%%CITATION = HEP-TH 0206103;%%.

\bibitem{Bena:2003wd}
I.~Bena, J.~Polchinski, and R.~Roiban, ``{Hidden symmetries of the
{$\ads_5
  \times S^5$}superstring},'' {\em Phys. Rev.} {\bf D69} (2004) 046002,
\href{http://www.arXiv.org/abs/hep-th/0305116}{{\tt
hep-th/0305116}}.
%%CITATION = HEP-TH/0305116;%%.

\bibitem{Serban:2004jf}
D.~Serban and M.~Staudacher, ``{Planar {$\CN = 4$} gauge theory and
the
  Inozemtsev long range spin chain},'' {\em JHEP} {\bf 06} (2004) 001,
\href{http://www.arXiv.org/abs/hep-th/0401057}{{\tt
hep-th/0401057}}.
%%CITATION = HEP-TH/0401057;%%.

\bibitem{Callan:2003xr}
J.~Callan, Curtis~G. {\em et al.}, ``Quantizing string theory in
{$\mbox{AdS}_5
  \times S^5$}: Beyond the {pp-wave},'' {\em Nucl. Phys.} {\bf B673} (2003)
  3--40,
\href{http://www.arXiv.org/abs/hep-th/0307032}{{\tt
hep-th/0307032}}.
%%CITATION = HEP-TH 0307032;%%.

\bibitem{Arutyunov:2004vx}
G.~Arutyunov, S.~Frolov, and M.~Staudacher, ``Bethe ansatz for
quantum
  strings,'' {\em JHEP} {\bf 10} (2004) 016,
\href{http://www.arXiv.org/abs/hep-th/0406256}{{\tt
hep-th/0406256}}.
%%CITATION = HEP-TH 0406256;%%.

\bibitem{Beisert:2005tm}
N.~Beisert, ``The {$\mathfrak{su}(2|2)$} dynamic {S}-matrix,''
\href{http://www.arXiv.org/abs/hep-th/0511082}{{\tt
hep-th/0511082}}.
%%CITATION = HEP-TH 0511082;%%.

\bibitem{Beisert:2006ez}
N.~Beisert, B.~Eden, and M.~Staudacher, ``{Transcendentality and
crossing},''
  {\em J. Stat. Mech.} {\bf 0701} (2007) P021,
\href{http://www.arXiv.org/abs/hep-th/0610251}{{\tt
hep-th/0610251}}.
%%CITATION = HEP-TH/0610251;%%.

\bibitem{Harmark:2008gm}
T.~Harmark, K.~R. Kristjansson, and M.~Orselli, \href{http://arxiv.org/abs/0806.3370}{{\tt
arXiv:0806.3370 [hep-th]}}.
%%CITATION = 0806.3370;%%.
.

\bibitem{Harmark:2006di}
T.~Harmark and M.~Orselli, ``Quantum mechanical sectors in thermal
{$\CN = 4$}
  super {Yang-Mills} on {$\mathds{R} \times S^3$},'' {\em Nucl. Phys.} {\bf
  B757} (2006) 117--145,
\href{http://www.arXiv.org/abs/hep-th/0605234}{{\tt
hep-th/0605234}}.
%%CITATION = HEP-TH 0605234;%%.

\bibitem{Harmark:2006ta}
T.~Harmark and M.~Orselli, ``Matching the {Hagedorn} temperature in
  {AdS/CFT},'' {\em Phys. Rev.} {\bf D74} (2006) 126009,
\href{http://www.arXiv.org/abs/hep-th/0608115}{{\tt
hep-th/0608115}}.
%%CITATION = HEP-TH 0608115;%%.

\bibitem{Harmark:2006ie}
T.~Harmark, K.~R. Kristjansson, and M.~Orselli, ``Magnetic
{Heisenberg-chain} /
  pp-wave correspondence,'' {\em JHEP} {\bf 02} (2007) 085,
\href{http://www.arXiv.org/abs/hep-th/0611242}{{\tt
hep-th/0611242}}.
%%CITATION = HEP-TH/0611242;%%.

\bibitem{Harmark:2007px}
T.~Harmark, K.~R. Kristjansson, and M.~Orselli, ``Decoupling limits
of
  {$\CN=4$} super {Yang-Mills} on {$\R \times S^3$},''
\href{http://www.arXiv.org/abs/arXiv:0707.1621 [hep-th]}{{\tt
arXiv:0707.1621
  [hep-th]}}.
%%CITATION = ARXIV:0707.1621;%%.

\bibitem{Harmark:2007et}
T.~Harmark, K.~R. Kristjansson, and M.~Orselli, `` {The {Hagedorn} temperature
  in a decoupled sector of {AdS/CFT}},'' {\em Fortsch. Phys.} {\bf 55} (2007)
  754--759,
\href{http://www.arXiv.org/abs/hep-th/0701088}{{\tt
hep-th/0701088}}.
%%CITATION = HEP-TH/0701088;%%.

\bibitem{Beisert:2003xu}
N.~Beisert, J.~A. Minahan, M.~Staudacher, and K.~Zarembo,
``Stringing spins and
  spinning strings,'' {\em JHEP} {\bf 09} (2003) 010,
\href{http://www.arXiv.org/abs/hep-th/0306139}{{\tt
hep-th/0306139}}.
%%CITATION = HEP-TH 0306139;%%.

\bibitem{Arutyunov:2003uj}
G.~Arutyunov, S.~Frolov, J.~Russo, and A.~A. Tseytlin, ``{Spinning
strings in
  {$\ads_5 \times S^5$} and integrable systems},'' {\em Nucl. Phys.} {\bf B671}
  (2003) 3--50,
\href{http://www.arXiv.org/abs/hep-th/0307191}{{\tt
hep-th/0307191}}.
%%CITATION = HEP-TH/0307191;%%.

\bibitem{Minahan:2005mx}
J.~A. Minahan, A.~Tirziu, and A.~A. Tseytlin, ``{$1/J$} corrections
to
  semiclassical {AdS/CFT} states from quantum {Landau-Lifshitz} model,'' {\em
  Nucl. Phys.} {\bf B735} (2006) 127--171,
\href{http://www.arXiv.org/abs/hep-th/0509071}{{\tt
hep-th/0509071}}.
%%CITATION = HEP-TH/0509071;%%.

\bibitem{Minahan:2005qj}
J.~A. Minahan, A.~Tirziu, and A.~A. Tseytlin, ``{$1/J^2$}
corrections to {BMN}
  energies from the quantum long range {Landau-Lifshitz} model,'' {\em JHEP}
  {\bf 11} (2005) 031,
\href{http://www.arXiv.org/abs/hep-th/0510080}{{\tt
hep-th/0510080}}.
%%CITATION = HEP-TH/0510080;%%.

\bibitem{Fradkin}
{See for example chapter 5 in: E. Fradkin}, ``{Field theories of
condensed
  matter systems},'' {\em Addison-Wesley Publishing Company, Redwood City, CA}
  (1991).

\bibitem{Kruczenski:2003gt}
M.~Kruczenski, ``Spin chains and string theory,'' {\em Phys. Rev.
Lett.} {\bf
  93} (2004) 161602,
\href{http://www.arXiv.org/abs/hep-th/0311203}{{\tt
hep-th/0311203}}.
%%CITATION = HEP-TH 0311203;%%.

\bibitem{Kruczenski:2004kw}
M.~Kruczenski, A.~V. Ryzhov, and A.~A. Tseytlin, ``{Large spin limit
of
  {$\ads_5\times S^5$} string theory and low energy expansion of ferromagnetic
  spin chains},'' {\em Nucl. Phys.} {\bf B692} (2004) 3--49,
\href{http://www.arXiv.org/abs/hep-th/0403120}{{\tt
hep-th/0403120}}.
%%CITATION = HEP-TH/0403120;%%.

\bibitem{Beisert:2005mq}
N.~Beisert, A.~A. Tseytlin, and K.~Zarembo, ``Matching quantum
strings to
  quantum spins: One-loop vs. finite-size corrections,'' {\em Nucl. Phys.} {\bf
  B715} (2005) 190--210,
\href{http://www.arXiv.org/abs/hep-th/0502173}{{\tt
hep-th/0502173}}.
%%CITATION = HEP-TH 0502173;%%.

\bibitem{Hernandez:2005nf}
R.~Hernandez, E.~Lopez, A.~Perianez, and G.~Sierra, ``{Finite size
effects in
  ferromagnetic spin chains and quantum corrections to classical strings},''
  {\em JHEP} {\bf 06} (2005) 011,
\href{http://www.arXiv.org/abs/hep-th/0502188}{{\tt
hep-th/0502188}}.
%%CITATION = HEP-TH/0502188;%%.

\bibitem{Ahmed:1978uw}
S.~Ahmed, M.~Bruschi, F.~Calogero, M.~A. Olshanetsky, and A.~M.
Perelomov,
  ``Properties of the zeros of the classical polynomials and of {Bessel}
  functions,''. ROME-110-1978.

\bibitem{Arutyunov:2003za}
G.~Arutyunov, J.~Russo, and A.~A. Tseytlin, ``{Spinning strings in
  {$\ads_5\times S^5$}: New integrable system relations},'' {\em Phys. Rev.}
  {\bf D69} (2004) 086009,
\href{http://www.arXiv.org/abs/hep-th/0311004}{{\tt
hep-th/0311004}}.
%%CITATION = HEP-TH/0311004;%%.

%\cite{Gromov:2005gp}
\bibitem{Gromov:2005gp}
  N.~Gromov and V.~Kazakov,
  ``Double scaling and finite size corrections in $\mathfrak{sl}(2)$ spin chain,''
  Nucl.\ Phys.\  B {\bf 736}, 199 (2006),
  \href{http://www.arXiv.org/abs/hep-th/0510194}{{\tt
hep-th/0510194}}.
  %%CITATION = NUPHA,B736,199;%%

\bibitem{Gromov:2007ky}
  N.~Gromov and P.~Vieira,
  ``Complete 1-loop test of AdS/CFT,''
  JHEP {\bf 0804}, 046 (2008),
  \href{http://www.arXiv.org/abs/arXiv:0709.3487 [hep-th]}{{\tt
arXiv:0709.3487
  [hep-th]}}.
  %%CITATION = JHEPA,0804,046;%%.

\bibitem{Bertolini:2002nr}
M.~Bertolini, J.~de~Boer, T.~Harmark, E.~Imeroni, and N.~A. Obers,
``Gauge
  theory description of compactified pp-waves,'' {\em JHEP} {\bf 01} (2003)
  016,
\href{http://www.arXiv.org/abs/hep-th/0209201}{{\tt
hep-th/0209201}}.
%%CITATION = HEP-TH 0209201;%%.

\bibitem{Ideguchi:2004wm}
K.~Ideguchi, ``Semiclassical strings on {$\ads_5 \times S^5/Z_M$}
and operators
  in orbifold field theories,'' {\em JHEP} {\bf 09} (2004) 008,
\href{http://www.arXiv.org/abs/hep-th/0408014}{{\tt
hep-th/0408014}}.
%%CITATION = HEP-TH/0408014;%%.

\bibitem{Beisert:2005he}
N.~Beisert and R.~Roiban, ``The {Bethe} ansatz for {$Z_S$} orbifolds
of {$\CN =
  4$} super {Yang- Mills} theory,'' {\em JHEP} {\bf 11} (2005) 037,
\href{http://www.arXiv.org/abs/hep-th/0510209}{{\tt
hep-th/0510209}}.
%%CITATION = HEP-TH 0510209;%%.

\bibitem{Astolfi:2006is}
D.~Astolfi, V.~Forini, G.~Grignani, and G.~W. Semenoff, ``Finite
size
  corrections and integrability of {$\CN = 2$} {SYM} and {DLCQ} strings on a
  pp-wave,'' {\em JHEP} {\bf 09} (2006) 056,
\href{http://www.arXiv.org/abs/hep-th/0606193}{{\tt
hep-th/0606193}}.
%%CITATION = HEP-TH/0606193;%%.

\bibitem{Larsen:2007bm}
K.~J. Larsen and N.~A. Obers, ``Phases of thermal {$\CN=2$} quiver
gauge
  theories,''
\href{http://www.arXiv.org/abs/arXiv:0708.3199 [hep-th]}{{\tt
arXiv:0708.3199
  [hep-th]}}.
%%CITATION = ARXIV:0708.3199;%%.

\bibitem{windingstring}
D.~Astolfi, G.~Grignani, T.~Harmark, and M.~Orselli, ``in
preparation,''.

\bibitem{Szabo:2001kg}
R.~J. Szabo, ``{Quantum field theory on noncommutative spaces},''
{\em Phys.
  Rept.} {\bf 378} (2003) 207--299,
\href{http://www.arXiv.org/abs/hep-th/0109162}{{\tt
hep-th/0109162}}.
%%CITATION = HEP-TH/0109162;%%.

\bibitem{Lubcke:2004dg}
M.~Lubcke and K.~Zarembo, ``Finite-size corrections to anomalous
dimensions in
  {$\CN = 4$} {SYM} theory,'' {\em JHEP} {\bf 05} (2004) 049,
\href{http://www.arXiv.org/abs/hep-th/0405055}{{\tt
hep-th/0405055}}.
%%CITATION = HEP-TH/0405055;%%.

\bibitem{Bargheer:2008kj}
T.~Bargheer, N.~Beisert, and N.~Gromov, ``{Quantum Stability for the
Heisenberg
  Ferromagnet},''
\href{http://www.arXiv.org/abs/0804.0324}{{\tt 0804.0324}}.
%%CITATION = 0804.0324;%%.

\bibitem{Stefanski:2004cw}
J.~Stefanski, B. and A.~A. Tseytlin, ``Large spin limits of
{AdS/CFT} and
  generalized {Landau-Lifshitz} equations,'' {\em JHEP} {\bf 05} (2004) 042,
\href{http://www.arXiv.org/abs/hep-th/0404133}{{\tt
hep-th/0404133}}.
%%CITATION = HEP-TH 0404133;%%.

\bibitem{Bellucci:2004qr}
S.~Bellucci, P.~Y. Casteill, J.~F. Morales, and C.~Sochichiu,
``{$SL(2)$} spin
  chain and spinning strings on {$\mbox{AdS}_5 \times S^5$},'' {\em Nucl.
  Phys.} {\bf B707} (2005) 303--320,
\href{http://www.arXiv.org/abs/hep-th/0409086}{{\tt
hep-th/0409086}}.
%%CITATION = HEP-TH 0409086;%%.

\bibitem{SchaferNameki:2005tn}
S.~Schafer-Nameki, M.~Zamaklar, and K.~Zarembo, ``{Quantum
corrections to
  spinning strings in {$\ads_5\times S^5$} and Bethe ansatz: A comparative
  study},'' {\em JHEP} {\bf 09} (2005) 051,
\href{http://www.arXiv.org/abs/hep-th/0507189}{{\tt
hep-th/0507189}}.
%%CITATION = HEP-TH/0507189;%%.

\bibitem{Frolov:2003tu}
S.~Frolov and A.~A. Tseytlin, ``{Quantizing three-spin string
solution in
  {$\ads_5 \times S^5$}},'' {\em JHEP} {\bf 07} (2003) 016,
\href{http://www.arXiv.org/abs/hep-th/0306130}{{\tt
hep-th/0306130}}.
%%CITATION = HEP-TH/0306130;%%.

\bibitem{Frolov:2004bh}
S.~A. Frolov, I.~Y. Park, and A.~A. Tseytlin, ``{On one-loop
correction to
  energy of spinning strings in {$S^5$}},'' {\em Phys. Rev.} {\bf D71} (2005)
  026006,
\href{http://www.arXiv.org/abs/hep-th/0408187}{{\tt
hep-th/0408187}}.
%%CITATION = HEP-TH/0408187;%%.

\bibitem{Park:2005ji}
I.~Y. Park, A.~Tirziu, and A.~A. Tseytlin, ``Spinning strings in
{$\mbox{AdS}_5
  \times S^5$}: One-loop correction to energy in $sl(2)$ sector,'' {\em JHEP}
  {\bf 03} (2005) 013,
\href{http://www.arXiv.org/abs/hep-th/0501203}{{\tt
hep-th/0501203}}.
%%CITATION = HEP-TH 0501203;%%.

\bibitem{Beisert:2005bv}
N.~Beisert and L.~Freyhult, ``Fluctuations and energy shifts in the
{Bethe}
  ansatz,'' {\em Phys. Lett.} {\bf B622} (2005) 343--348,
\href{http://www.arXiv.org/abs/hep-th/0506243}{{\tt
hep-th/0506243}}.
%%CITATION = HEP-TH 0506243;%%.

\bibitem{Grignani:2005yv}
G.~Grignani, M.~Orselli, B.~Ramadanovic, G.~W. Semenoff, and
D.~Young,
  ``Divergence cancellation and loop corrections in string field theory on a
  plane wave background,'' {\em JHEP} {\bf 12} (2005) 017,
\href{http://www.arXiv.org/abs/hep-th/0508126}{{\tt
hep-th/0508126}}.
%%CITATION = HEP-TH/0508126;%%.

\bibitem{Grignani:2006en}
G.~Grignani, M.~Orselli, B.~Ramadanovic, G.~W. Semenoff, and
D.~Young,
  ``{AdS/CFT vs. string loops},'' {\em JHEP} {\bf 06} (2006) 040,
\href{http://www.arXiv.org/abs/hep-th/0605080}{{\tt
hep-th/0605080}}.
%%CITATION = HEP-TH/0605080;%%.

\bibitem{Douglas:1996sw}
M.~R. Douglas and G.~W. Moore, ``{D-branes, Quivers, and {ALE}
Instantons},'' \href{http://www.arXiv.org/abs/hep-th/9603167}{{\tt
hep-th/9603167}}.
%%CITATION = HEP-TH/9603167;%%.

\bibitem{Beisert:2002ff}
N.~Beisert, C.~Kristjansen, J.~Plefka, and M.~Staudacher, ``{BMN}
gauge theory
  as a quantum mechanical system,'' {\em Phys. Lett.} {\bf B558} (2003)
  229--237,
\href{http://www.arXiv.org/abs/hep-th/0212269}{{\tt
hep-th/0212269}}.
%%CITATION = HEP-TH 0212269;%%.

\bibitem{Beisert:2003jb}
N.~Beisert, ``Higher loops, integrability and the near {BMN}
limit,'' {\em
  JHEP} {\bf 09} (2003) 062,
\href{http://www.arXiv.org/abs/hep-th/0308074}{{\tt
hep-th/0308074}}.
%%CITATION = HEP-TH 0308074;%%.

\bibitem{Eden:2004ua}
B.~Eden, C.~Jarczak, and E.~Sokatchev, ``A three-loop test of the
dilatation
  operator in {$\CN = 4$} {SYM},'' {\em Nucl. Phys.} {\bf B712} (2005)
  157--195,
\href{http://www.arXiv.org/abs/hep-th/0409009}{{\tt
hep-th/0409009}}.
%%CITATION = HEP-TH/0409009;%%.

\bibitem{DeRisi:2004bc}
G.~De~Risi, G.~Grignani, M.~Orselli, and G.~W. Semenoff, ``{DLCQ}
string
  spectrum from {$\CN = 2$} {SYM} theory,'' {\em JHEP} {\bf 11} (2004) 053,
\href{http://www.arXiv.org/abs/hep-th/0409315}{{\tt
hep-th/0409315}}.
%%CITATION = HEP-TH/0409315;%%.

\bibitem{Lunin:2005jy}
O.~Lunin and J.~M. Maldacena, ``{Deforming field theories with
{$U(1) \times
  U(1)$} global symmetry and their gravity duals},'' {\em JHEP} {\bf 05} (2005)
  033,
\href{http://www.arXiv.org/abs/hep-th/0502086}{{\tt
hep-th/0502086}}.
%%CITATION = HEP-TH/0502086;%%.

\bibitem{Frolov:2005ty}
S.~A. Frolov, R.~Roiban, and A.~A. Tseytlin, ``{Gauge - string
duality for
  superconformal deformations of {$\CN = 4$} super {Yang-Mills} theory},'' {\em
  JHEP} {\bf 07} (2005) 045,
\href{http://www.arXiv.org/abs/hep-th/0503192}{{\tt
hep-th/0503192}}.
%%CITATION = HEP-TH/0503192;%%.

\end{thebibliography}

\end{document}